\documentclass[pdflatex,sn-mathphys]{sn-jnl}

\jyear{2021}
\theoremstyle{thmstyleone}

\theoremstyle{thmstyletwo}%
\theoremstyle{thmstylethree}%
\usepackage{lmodern}
\usepackage{graphicx}
\usepackage{amsmath}
\usepackage{amssymb}
\usepackage{multirow}
\usepackage{longtable}
\usepackage{array}
\usepackage{makecell}
\usepackage{mathrsfs}
\usepackage[T1]{fontenc}
\NewExpandableDocumentCommand\XGap{m}{\noalign{\vskip #1}}
\NewExpandableDocumentCommand\Gap{}{\XGap{3pt}}
\newif\iffinal
\finaltrue 

\iffinal
\else
    \usepackage{xcolor}
    \usepackage[normalem]{ulem}
     
    \newcommand\deleted{\bgroup\markoverwith{\textcolor{blue}{\rule[0.5ex]{2pt}{0.8pt}}}\ULon}
\fi

\iffinal
    \newcommand\deleted[1]{}    
\fi

\begin{document}

\title[PassionNet]{PassionNet: An Innovative Framework for Duplicate and Conflicting Requirements Identification}

 \author*[1,2]{\fnm{Summra} \sur{Saleem}}\email{summra.saleem@dfki.de}
\equalcont{These authors contributed equally to this work.}

\author*[2]{\fnm{Muhammad Nabeel} \sur{Asim}}\email{Muhammad\_Nabeel.Asim@dfki.de}
 \equalcont{These authors contributed equally to this work.}

\author[1,2]{\fnm{Andreas} \sur{Dengel}}\email{andreas.dengel@dfki.de}

 \affil[1]{\orgdiv{Department of Computer Science}, \orgname{Rhineland-Palatinate Technical University of Kaiserslautern-Landau}, \orgaddress{\city{Kaiserslautern}, \postcode{67663},  \country{Germany}}}
 
\affil[2]{\orgname{German Research Center for Artificial Intelligence GmbH}, \orgaddress{\city{Kaiserslautern}, \postcode{67663},  \country{Germany}}}

\abstract{ Early detection and resolution of duplicate and conflicting requirements can significantly enhance project efficiency and overall software quality. Researchers have developed various computational predictors by leveraging Artificial Intelligence (AI) potential to detect duplicate and conflicting requirements. However, these predictors lack in performance and requires more effective approaches to empower software development processes. Following the need of a unique predictor that can accurately identify duplicate and conflicting requirements, this research offers a comprehensive framework that facilitate development of 3 different types of predictive pipelines: language models based, multi-model similarity knowledge-driven and large language models (LLMs) context + multi-model similarity knowledge-driven. Within first type predictive pipelines landscape, framework facilitates conflicting/duplicate requirements identification by leveraging 8 distinct types of LLMs. In second type, framework supports development of predictive pipelines that leverage multi-scale and multi-model similarity knowledge, ranging from traditional similarity computation methods to advanced similarity vectors generated by LLMs. In the third type, the framework synthesizes predictive pipelines by integrating contextual insights from LLMs with multi-model similarity knowledge. Across 6 public benchmark datasets, extensive testing of 760 distinct predictive pipelines including 8 standalone LLM-based, 112 similarity knowledge-Driven, and 640 hybrid LLM contextual and similarity Knowledge-Driven demonstrates that  hybrid predictive pipelines consistently outperforms other two types predictive pipelines in accurately identifying duplicate and conflicting requirements. This predictive pipeline outperformed existing state-of-the-art predictors performance with an overall performance margin of 13\% in terms of F1-score.}


\maketitle
\section{Introduction}
Requirements engineering (RE) is a key pillar of software development life cycle (SDLC) models, as it guides development teams through the critical stages of planning, implementation, testing and deployment \citep{mohapatra2020fundamentals, akbar2020systematic}. It ensures that software solutions effectively meet stakeholder needs and contributes to the overall quality, efficiency, and success of software projects \citep{pargaonkar2023comprehensive, saleem7mlr}. Despite its importance, RE is inherently complex due to the involvement of diverse stakeholders with varying levels of technical expertise \citep{heyn2021requirement}. This Interdisciplinary nature of stakeholders often results in significant communication gaps between technical experts (such as developers and system architects) and non-technical participants (including end-users, business analysts, and management) \citep{pereira2021towards}. The communication gap often leads to duplicate requirements, where the same functionality is described differently \citep{motger2020resim}. For instance, one requirement might state, ``The system shall be able to process payments" while another requirement may contain same information but in different words as ``The software must support online payment transactions \citep{chazette2020explainability}." Duplicate requirements lead to unintentional redundancy in  implementation efforts such as development team may allocate valuable time and resources to independently implementing the same functionality in different parts of the project. it leads towards inefficiencies, increased costs, and delayed project timelines \citep{saleem7mlr, saleem2023fnreq}.

Within Requirement engineering landscape another challenge is conflicting requirements where two or more requirements introduce incompatible or
contradictory software functionalities \citep{ren2024requirements}. For instance, one requirement might contain payment processing information of one type such as,  ``The system shall be able to process payments within 24 hours \citep{saarni2021check}." While another requirement may contain  payment processing information of another type as ``All payment transactions must be processed instantly." Such conflicting requirements can lead to project delays, increased costs and potential system failure by creating ambiguity and misalignment among stakeholders \citep{scientific2024detectiing}. These conflicts can lead to re-iterations of project phases, extra effort expended, and ultimately compromise the successful development and implementation of software systems \citep{malik2023automated}. 

Conventionally, both types duplicate and conflicting requirements are identified manually \citep{guo2021automatically}. As software complexity increase, requirements size also grow exponentially which makes manual identification inefficient and error-prone \citep{malik2023transfer}. To overcome this challenge, and expedite the identification of conflicting and duplicate requirements, researchers are developing AI-driven applications by leveraging the potential of Artificial Intelligence approaches \citep{motger2020resim, falessi2010comprehensive, rao2022research, gambo2024identifying, gartner2024automated}. AI-driven applications are developed by making all possible pairs of requirements \citep{helmeczi2023few}. Requirement pairs containing duplicate content are labeled as ``duplicate," while those without duplicate content are labeled as ``neutral" \citep{malik2023data}. Similarly, conflicting requirements identification task, requirement pairs are labeled with two distinct classes: ``conflict" for pairs containing contradictory content and ``neutral" for pairs that do not have contradictory information. During the training phase, AI models analyze requirement pairs and their associated class labels to learn patterns and relationships between  requirement pairs and labels. In the inference stage, the trained models makes use learned patterns to predict new or unseen requirement pairs corresponding class labels. AI field has witnessed remarkable advancements in strategies designed to extract and learn meaningful patterns and relationships between textual data and their corresponding labels. The potential of various AI strategies, including Convolutional Neural Networks (CNNs) \citep{xuan2024tool, wang2023deep}, Long Short-Term Memory networks (LSTMs) \citep{wang2023deep}, and language models \citep{abeba2022identification, malik2023automated}, has been explored for development of duplicate and conflicting requirements identification predictors. However, these predictors could not manage to produce optimal performance, primarily due to their inability to effectively extract and learn meaningful patterns within requirement pairs. This limitation hinders their ability to accurately classify pairs as duplicate or neutral and conflicting or neutral requirements.

To address these limitations, this study proposes to integrate requirement pairs similarity information within AI-predictive pipelines. This integration enables predictive pipelines to harnesses the deep semantic understanding of LLMs with invaluable insights derived from similarity knowledge and discriminative capabilities of ML classifiers. To thoroughly evaluate the effectiveness of similarity knowledge  integration into AI-predictive pipelines, this study introduces a comprehensive framework. Specifically, proposed framework offers development of  3 distinct types of predictive pipelines including; 1) LLMs context based, 2) multimodel similarity knowledge driven and 3) LLMS context + multimodel similarity knowledge-driven. The primary contributions of this research are summarized as follows:
\begin{itemize}
     \item  Across 6 requirements duplication and contradiction detection datasets,  it benchmarks performance of 8 distinct types of LLMs including: ALBERT \citep{lan2019albert}, BERT \citep{devlin2018bert}, DeBERTa \citep{he2020deberta}, Electra \citep{clark2020electra} GPT \citep{radford2019language}, Longformer \citep{beltagy2020longformer}, RoBERTa \citep{liu2019roberta} and XLNet \citep{yang2019xlnet}.
    \item It evaluates performance of 112 distinct multimodel similarity knowledge driven predictive pipelines, developed from 2 traditional representation learning methods (TFIDF and OkapiBM25), 10 traditional similarity computation methods, 10 distinct LLM similarity scores and 16 different ML classifiers.
    \item It assesses performance of 640 different context aware and multi-scale multimodel similarity-driven predictive pipelines that are developed from combined potential of LLMs contextual information,  multimodel similarity knowledge and 16 ML classifiers.  
    \item It provide in-depth performance comparison of top-performing  predictive pipelines of each category: 1) language models context 2) multimodel similarity knowledge and 3) LLMs context + multimodel similarity knowledge.
    \item It compares performance of framework top performing predictive pipelines and existing state-of-the-art predictors across 6 public datasets of conflict/duplicate requirement identification.
\end{itemize}

\section{Related Work}
This section offers valuable insights into $32$ predictors \citep{reddivari2022calculating, oktaviyani2020redundancy, natt2002feasibility, sillaber2013improving, motger2020resim, falessi2010comprehensive, rao2022research, abualhaija2024replication, scientific2024detectiing, binamungu2018detecting, rago2016identifying, luciv2018detecting, guo2021automatically, elhassan2022requirements, shah2021detecting, malik2023automated, moser2011requirements, liu2012applying, haffner2023introducing, abeba2022identification, helmeczi2022prompt, gambo2024identifying, gartner2024automated, malik2022supervised, kisso2024requirements, xuan2024tool, wang2023deep, liu2016cdnfre, liu2010cdade, malik2023transfer, malik2023data, helmeczi2023few} developed over past $10$ years for conflict/duplicate detection in requirements. Among these, specifically 12 predictors \citep{reddivari2022calculating, oktaviyani2020redundancy, natt2002feasibility, sillaber2013improving, motger2020resim, falessi2010comprehensive, rao2022research, abualhaija2024replication, scientific2024detectiing, binamungu2018detecting, rago2016identifying, luciv2018detecting} focus on duplicate detection, while 17 predictors \citep{guo2021automatically, elhassan2022requirements, shah2021detecting, malik2023automated, moser2011requirements, liu2012applying, haffner2023introducing, abeba2022identification, helmeczi2022prompt, gambo2024identifying, gartner2024automated, malik2022supervised, kisso2024requirements, xuan2024tool, wang2023deep, liu2016cdnfre, liu2010cdade} target conflict identification. However,  only 3 \citep{malik2023transfer, malik2023data, helmeczi2023few} predictors have been proposed  to simultaneously address both duplicate and conflict detection among requirements. Moreover, based on working paradigms, these predictors fall into 7 different categorizes: 1) Rule-based \citep{guo2021automatically}, 2) Statistical \citep{reddivari2022calculating, oktaviyani2020redundancy, natt2002feasibility, sillaber2013improving, elhassan2022requirements, shah2021detecting, malik2023automated}, 3) Ontology based \citep{moser2011requirements, liu2012applying} 4) Machine learning based \citep{motger2020resim, falessi2010comprehensive}, 5) Deep learning based \citep{haffner2023introducing, abeba2022identification}, 6)
Language models based \citep{rao2022research, malik2023transfer, malik2023data, helmeczi2023few, helmeczi2022prompt, gambo2024identifying, gartner2024automated, malik2022supervised, kisso2024requirements, xuan2024tool, wang2023deep}, 7) Tools based \citep{abualhaija2024replication, scientific2024detectiing, binamungu2018detecting, rago2016identifying, luciv2018detecting, liu2016cdnfre, liu2010cdade} predictors. 

Guo et al. \citep{guo2021automatically} designed a rule-based predictor by using finer semantics and heuristic rules for requirements conflict detection. However, it \citep{guo2021automatically} lacks flexibility to adopt for unseen data, which makes it unsuitable for frequently evolving nature of requirements. Among 8 statistical algorithm based predictors \citep{reddivari2022calculating, oktaviyani2020redundancy, natt2002feasibility, sillaber2013improving, elhassan2022requirements, shah2021detecting, malik2023automated}, 4 predictors \citep{sillaber2013improving, elhassan2022requirements, shah2021detecting, malik2023automated} utilized clustering algorithm including, mean shift and correlative clustering. On the other hand, remaining 4 predictors \citep{reddivari2022calculating, oktaviyani2020redundancy, natt2002feasibility, sillaber2013improving} used similarity based approaches. Specifically one predictor \citep{reddivari2022calculating} explored combined potential of term frequency inverse document frequency (TFIDF) and cosine similarity algorithm, while one predictor \citep{oktaviyani2020redundancy} leveraged fast heuristic similarity, and two predictors \citep{malik2023automated, natt2002feasibility} made use of dice and cosine similarity. 

In existing literature, two predictors \citep{moser2011requirements, liu2012applying} utilized ontological model and predefined rule to address conflict requirement detection. Specifically, these predictors are designed using domain knowledge to offer more formal and explicit representation of shared conceptualization. Besides ontology-based predictors, two machine learning based predictors \citep{motger2020resim, falessi2010comprehensive} utilized TFIDF with LogR \citep{falessi2010comprehensive} and SVM \citep{motger2020resim} classifiers. Although these predictors \citep{motger2020resim, falessi2010comprehensive} offer data driven approach for duplicate requirements detection and lack to capture intricate relationships among redundant features. Following the success of deep learning techniques in various natural language processing (NLP) task, two deep learning predictors \citep{haffner2023introducing, abeba2022identification} are developed. Specifically, one predictor \citep{haffner2023introducing} explored combined potential of document frequency and  bi-gram embedding with feed-forward neural network (FNN). Contrarily, other predictor \citep{abeba2022identification} utilized Word2Vec embedding methods to generate statistical vector along with Bi-LSTM classifier. 

The advent of language models has revolutionized NLP, leading to the development of LM-based predictors. These models can be adapted for downstream tasks through two primary approaches: 1) fine-tuning with a self-classifier, where the model's own output layer is adjusted for the specific task, and 2) fine-tuning using an external classifier, where the model's outputs are fed to separate classifier. In the field of requirements engineering, $4$ predictors have employed self-classifier approach, while $7$ predictors have leveraged external classifiers for conflict and duplicate detection. 4 LM based predictors \citep{malik2023transfer, malik2023data, helmeczi2023few} utilized  language models (BERT \citep{malik2023transfer}, DistilBERT \citep{malik2023data}, RoBERTA \citep{helmeczi2022prompt}, DeBERTa \citep{helmeczi2023few}) with self classifier, to simultaneously address both duplicate and conflict detection. The remaining 7 LM based predictors \citep{rao2022research, gambo2024identifying, gartner2024automated, malik2022supervised, kisso2024requirements, xuan2024tool, wang2023deep} explored diverse strategies. Specifically, 2 predictors \citep{gambo2024identifying, gartner2024automated} integrated formal and lexical logics with RoBERTa \citep{gambo2024identifying} and GPT-3 \citep{gartner2024automated}, one made use of BERT with cosine similarity \citep{kisso2024requirements}, one reaped benefits of both BERT and GPT-3.5 with cosine similarity \citep{kisso2024requirements}, and two incorporated ROBERTa \citep{xuan2024tool} and GPT-3 \citep{wang2023deep} with CNN classifiers for more comprehensive feature extraction. Lastly, one predictor \citep{rao2022research} explored potential of BERT with LSTM classifier for duplicate requirements detection. This versatility of language models to address conflict or duplicate requirements detection to, highlight rapid adaptation of advanced techniques in requirement engineering filed.

Lastly among $7$ tools based predictors, 5 predictors \citep{abualhaija2024replication, scientific2024detectiing, binamungu2018detecting, rago2016identifying, luciv2018detecting} namely; NLP4RE, KNIME, PMD/CPD, CloneDR, DECKARD, Mulan Toolkit, CloneMiner, focus on duplicate requirement detection. On the other hand, two predictors \citep{liu2016cdnfre, liu2010cdade} are developed for conflict requirement detection predictors including CDNFRE prototype \citep{liu2016cdnfre} and CDADE \citep{liu2010cdade} tools. Table \ref{related-work} illustrates comprehensive overview of existing conflict/duplicate requirement detection predictors in terms of dataset, domain, representation learning and predictor.

{\scriptsize\tabcolsep=2pt
\renewcommand{\arraystretch}{1.8}
\begin{longtable}{|p{3.cm}|p{3.5cm}|p{2.cm}|p{3cm}|}
\caption{Overview of existing conflict/duplicate requirement detection predictors}
\label{related-work}\\
\hline
\textbf{Author, Year [ref]} &
  \textbf{Dataset} &
  \textbf{Embedding} &
  \textbf{Predictor/Tool} \\ \hline
 \multicolumn{4}{|c|}{\textbf{Duplicate Requirement Detection}} \\ \hline
 Haija et al., 2024 \citep{abualhaija2024replication} &
  \_ &
  \_ &
  NLP4RE tool \\ \hline 
Raghad et al., 2024 \citep{scientific2024detectiing} &
 Common Utils  for Rapid Dataset, JSOUP, Junit &
  \_ &
  KNIME tool + DNN \\ \hline
Rao et al., 2022 \citep{rao2022research} &
  PURE &
  BERT &
  LSTM \\ \hline 
Reddivari et al., 2022 \citep{reddivari2022calculating} &
  PURE &
  TFIDF &
  Cosine Similarity \\ \hline
Motger et al., 2020 \citep{motger2020resim} &
  Qt's &
  TFIDF &
  SVM \\ \hline
Oktaviyani et al., 2020 \citep{oktaviyani2020redundancy} &
\_ &
  \_ &
  Fast Heuristic Similarity \\ \hline 
Binammungu et al., 2018 \citep{binamungu2018detecting} &
   jcshs, ATest,  Facad Services &
  \_ &
  PMD/CPD tool, CloneDR tool, DECKARD tool \\ \hline
Luciv et al., 2018 \citep{luciv2018detecting} &
  Luciv et al. Dataset &
  \_ &
  CloneMiner tool \\ \hline 
Rago et al., 2016 \citep{rago2016identifying} &
  DLibraCRM, MobileNews, WebJSARA, CRS, HWS &
 \_ &
  Mulan Toolkit \\ \hline
SIllaber et al., 2013 \citep{sillaber2013improving} &
  Incubator, ws, Jakarta &
  \_ &
  Clustering Algorithm \\ \hline
Faleesi et al., 2010 \citep{falessi2010comprehensive} &
  PROUD, ANTARCTICA, SELEX &
  TFIDF &
  LogR \\ \hline
Dag et al., 2002 \citep{natt2002feasibility} &
  Telelogic &
  \_ &
  Dice  \& Cosine Similarities \\ \hline
\multicolumn{4}{|c|}{\textbf{Conflicts Requirement Detection}} \\ \hline
Gambo et al., 2024 \citep{gambo2024identifying} &
   iOS App Store Dataset, Google Play Store Dataset &
   RoBERTa&
  RoBERTa \\ \hline
Gartner et al., 2024 \citep{gartner2024automated} &
  Dataset 1, Dataset 2, Dataset 3 &
  GPT-3 &
  GPT-3 \\ \hline
Kisso et al., 2024 \citep{kisso2024requirements} &
   Telecommunications, ATM, Video-on-demand,  PROMISE &
  SBERT + GPT-3.5 &
  Cosine Similarity \\ \hline
Xuan et al., 2024 \citep{xuan2024tool} &
  12 Datasets &
  RoBERTa &
  CNN \\ \hline
Wang et al., 2023 \citep{wang2023deep} &
  12 Datasets &
  RoBERTa &
  CNN \\ \hline
Haffner et al., 2023 \citep{haffner2023introducing} &
  ICG CrisisWatch reports &
IDF &
  FFN \\ \hline 
Malik et al., 2023 \citep{malik2023automated} &
  WorldVista,  Pure &
  TFIDF &
  Cosine Similarity \\ \hline
Shah et al., 2021 \citep{shah2021detecting} &
   Crime Tracking,  Fault-Tolerant Services,  Communication network, Tactical control system, e-Procurement &
  TFIDF &
  Correlative Clustering \\ \hline 
Guo et al., 2021 \citep{guo2021automatically} &
   UAV, WorldVista, Telecom Management, Solar Power &
 \_ &
  Heuristic Rules-based Algorithm \\ \hline
Abeba et al., 2022 \citep{abeba2022identification} &
  NFR &
  Word2Vec &
  Bi-LSTM \\ \hline
Elhassan et al., 2022 \citep{elhassan2022requirements} &
  \_ &
  \_ &
  Mean Shift Clustering Algorithm \\ \hline 
Malik et al., 2022 \citep{malik2022supervised} &
  OpenCoss, WorldVista, UAV, PURE, IBM-UAV &
  TFIDF, TFIDF + BERT, USE, BERT  &
  Phase 1: Cosine Similarity, Phase 2:  Entity Overlap \\ \hline
Helmeczi et al., 2022 \citep{helmeczi2022prompt} &
  Conflict Detection Dataset &
  RoBERTa &
  RoBERTa \\ \hline 
Liu et al., 2016 \citep{liu2016cdnfre} &
  NFR evolution metadata &
  \_ &
  CDNFREcprototype \\ \hline
Liu et al., 2012 \citep{liu2012applying} &
  \_ &
  \_ &
  NLP Tool \\ \hline 
Moser et al., 2011 \citep{moser2011requirements} &
  Trac Requiement Dataset &
  \_ &
  OntRep Tool \\ \hline 
Liu et al., 2010 \citep{liu2010cdade} & \_ &
  \_ &
  CDADE tool \\ \hline
\multicolumn{4}{|c|}{\textbf{Duplicate and Conflict Requirement Detection}} \\ \hline
Malik et al., 2023 \citep{malik2023transfer}  &
  UAV, WorldVista, PURE, CD/CNN &
  BERT &
  BERT  \\ \hline 
Malik et al., 2023 \citep{malik2023data} &
  WorldVista, UAV, PURE, OPENCOSS, StackOverflow, Bugzilla &
  BERT &
  BERT \\ \hline
Helmeczi et al., 2023 \citep{helmeczi2023few} &
  SRS conflict, Bugzilla, StackOverflow, Bugzilla &
  BERT &
  BERT \\ \hline  
\end{longtable}}
 
\section{Architectural Design of PassionNet Framework}
\begin{figure}
    \centering
    \centering
     \includegraphics[width=0.99\textwidth, height=2\textheight, keepaspectratio]{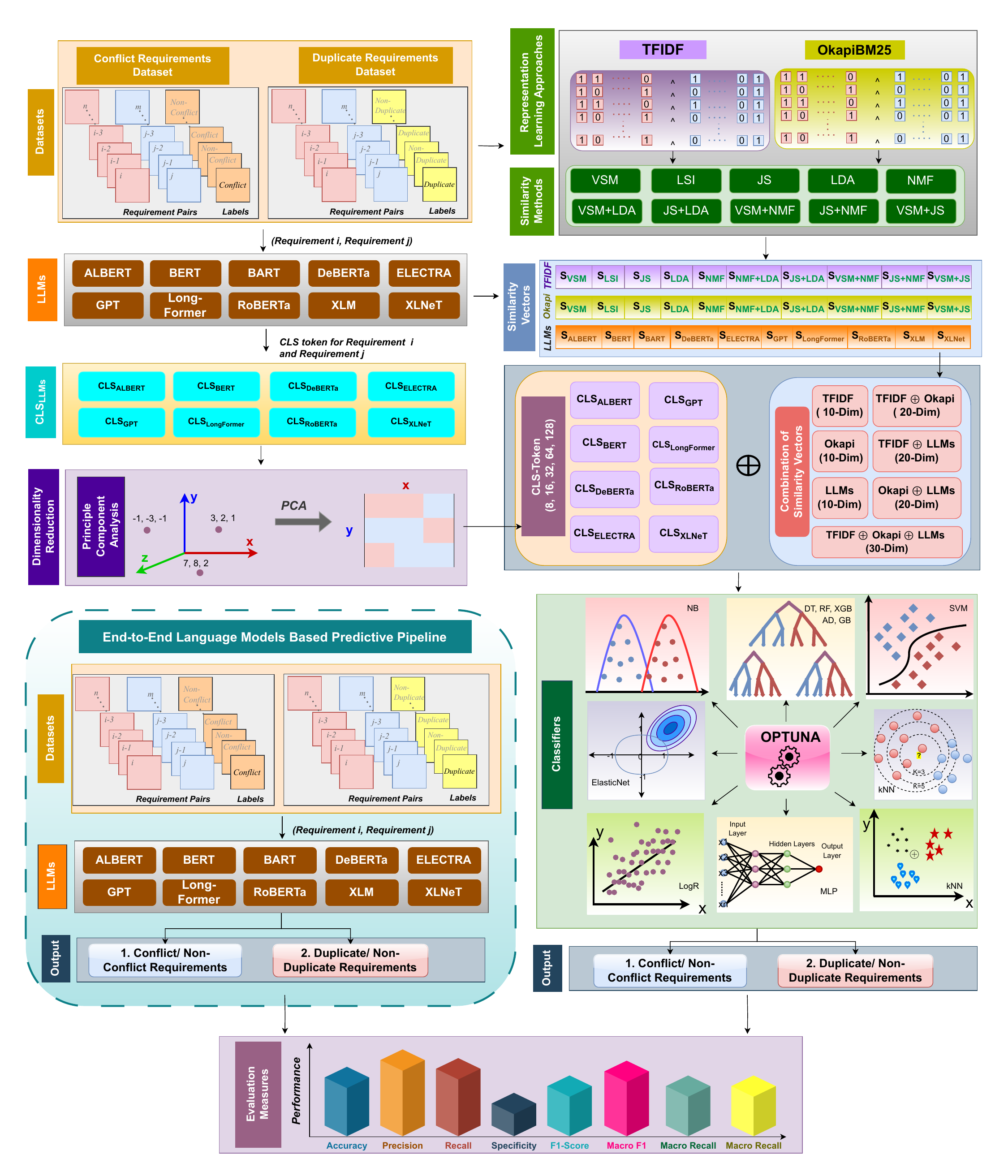}
    \caption{A High-Level Overview of PassionNet Framework}
    \label{Framework}
\end{figure}

Figure \ref{Framework} illustrates high-level overview of 3 distinct types of predictive pipelines that framework offers. In Figure \ref{Framework}, bottom left corner module with doted lines represents a pool of 8 distinct LLMs that offers development of first LLMs context aware predictive pipelines. The remaining modules of Figure \ref{Framework} illustrate a comprehensive overview of 752 distinct predictive pipelines, including 112 multimodel similarity knowledge driven pipelines (second type) and 640 LLM context + multimodel similarity knowledge driven predictive pipelines (third type). Following subsections illustrates details of all 3 types of predictive pipelines. 

\subsection{Large Language Models Context Aware Predictive Pipelines} 
The proposed framework offers diverse array of 8 different LLMs for conflict/duplicate requirements identification. Table \ref{LLMs-summary} provides overview of these LLMs in terms of architectures, layer configurations, masking technique, tokenization and attention mechanisms. The proposed framework leverages pre-trained language models that are subsequently refined through supervised fine-tuning to  capture deep linguistic and semantic relations for conflict/duplicate requirements identification. During fine-tuning, the requirements pairs are encapsulated into a single unit, beginning  with special [CLS] tokens  while two requirements are separated with [SEP] token. This format allows models to effectively process and capture semantic relations between requirement pairs.


\begin{table}[h!]
\resizebox{\textwidth}{!}{%
\begin{tabular}{lllllll}
\hline
\textbf{Model} & \textbf{Architecture} & \textbf{\begin{tabular}[c]{@{}l@{}}Number of\\ Layers in \\ Encoder\end{tabular}} & \textbf{\begin{tabular}[c]{@{}l@{}}Number of \\ Layers in \\ Decoder\end{tabular}} & \textbf{Tokenizer}          & \textbf{Attention Type}                                                                                 & \textbf{Masking Technique}                                                            \\ \hline
AlBert         & Encoder               & 12                                                                              & \_                                                                              & \begin{tabular}[c]{@{}l@{}}WordPiece/ \\SentencePiece\end{tabular} & Multi-head Self Attention                                                                               & Random                                                                                \\ \hline
Bert           & Encoder               & 24                                                                              & \_                                                                              & WordPiece                     & Multi-head Self Attention                                                                               & Random                                                                                \\ \hline
Deberta        & Encoder               & 24                                                                              & \begin{tabular}[c]{@{}l@{}}1 Enhanced \\ Mask  Decoder\end{tabular}             & WordPiece                     & Disentangled Attention                                                                                  & Random                                                                                \\ \hline
Electra        & Encoder               & 24                                                                              & \_                                                                              & WordPiece                     & Multi-head Self Attention                                                                               & \begin{tabular}[c]{@{}l@{}}Replaces tokens with\\ plausible alternatives\end{tabular} \\ \hline
Roberta        & Encoder               & 24                                                                              & \_                                                                              & BPE                           & Multi-head Self Attention                                                                               & Dynamic                                                                               \\ \hline
GPT            & Decoder               & \_                                                                              & 36                                                                              & BPE                           & \begin{tabular}[c]{@{}l@{}}Masked Multi-head \\ Self Attention\end{tabular}                             & \begin{tabular}[c]{@{}l@{}}Autoregressive\\ Approach\end{tabular}                     \\ \hline
Longformer     & \begin{tabular}[c]{@{}c@{}}Encoder-\\Decoder\end{tabular} & 6                                                                               & 6                                                                               & WordPiece                     & \begin{tabular}[c]{@{}l@{}}Local-context + \\ Global-context Attention\end{tabular}                     & Dynamic                                                                               \\ \hline
Roberta        & Encoder               & 12                                                                              & \_                                                                              & BPE                           & Multi-head Self Attention                                                                               & Dynamic                                                                               \\ \hline
xlnet          & Encoder               & 24                                                                              & \_                                                                              & SentencePiece                 & \begin{tabular}[c]{@{}l@{}}Self Attention (Content + Query)\\  For fine tuning drop Query\end{tabular}  & \begin{tabular}[c]{@{}l@{}}Permutation Based\\ Approach\end{tabular}                  \\ \hline
\end{tabular}
}
\caption{Overview of Language Models in terms of Architectures, Layer Configurations, Tokenization, Attention Mechanisms, and Masking}
\label{LLMs-summary}
\end{table}

\subsection{MultiModel Similarity Knowledge Driven Predictive Pipelines}\label{Similarity-Knowledge-informed-Predictive-pipelines}

This section illustrates the details of multimodel similarity knowledge driven predictive pipelines that comprises of 3 main modules: 1) traditional representation learning and similarity computation techniques, 2) LLMs based similarity computation and 3) classifiers.

\subsubsection{Traditional Similarity Techniques based Predictive Pipelines}
\begin{figure*}[htbp]
    \centering
    \begin{tabular}{cc}
        \includegraphics[width=0.45\textwidth]{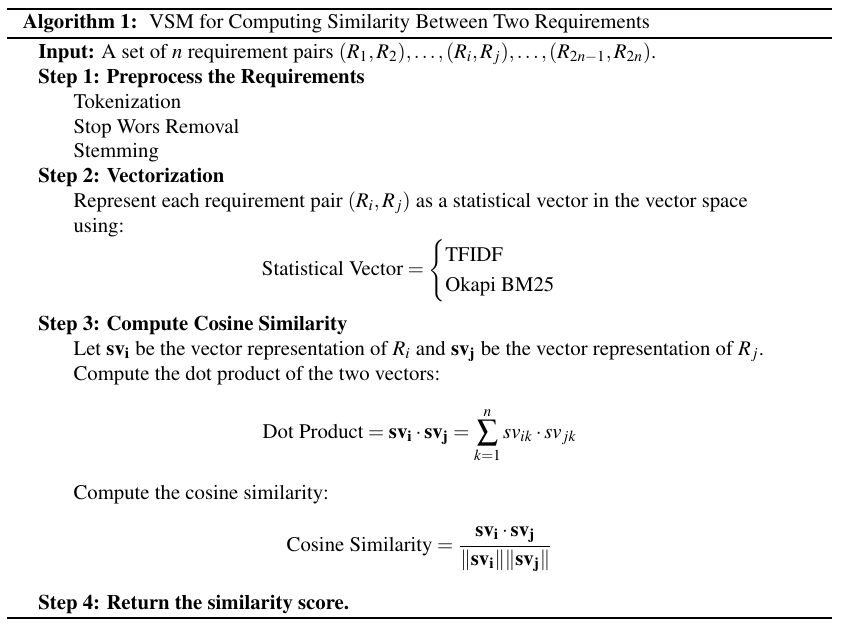} &
        \includegraphics[width=0.45\textwidth]{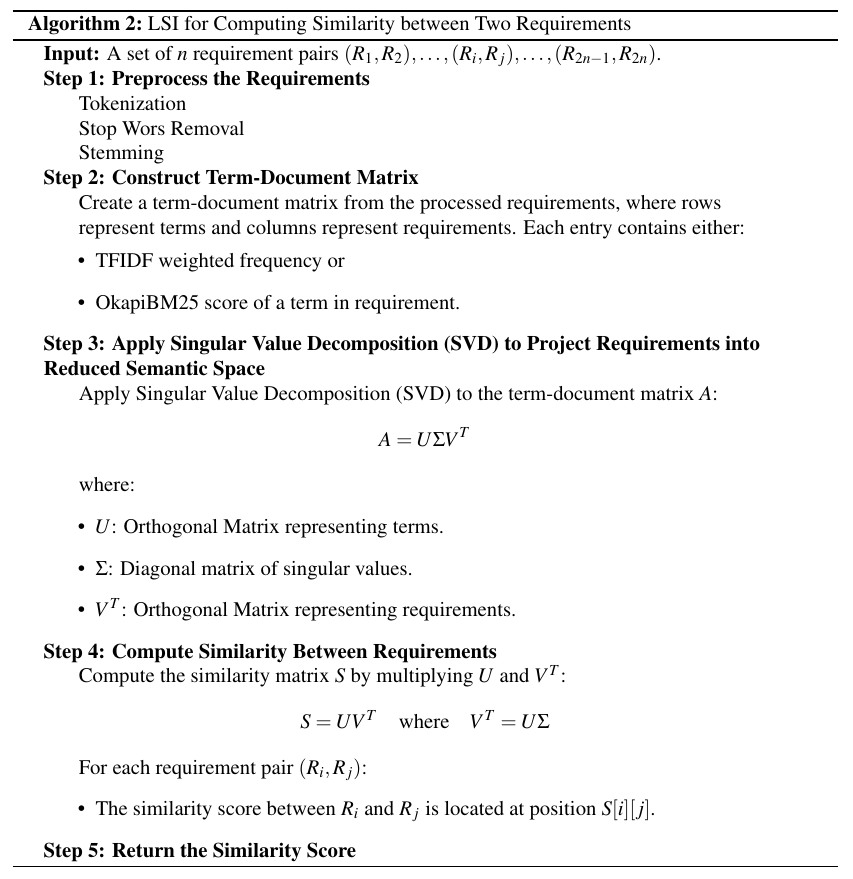} \\
        \textbf{VSM} & \textbf{LSI} \\[1pt]

        \includegraphics[width=0.45\textwidth]{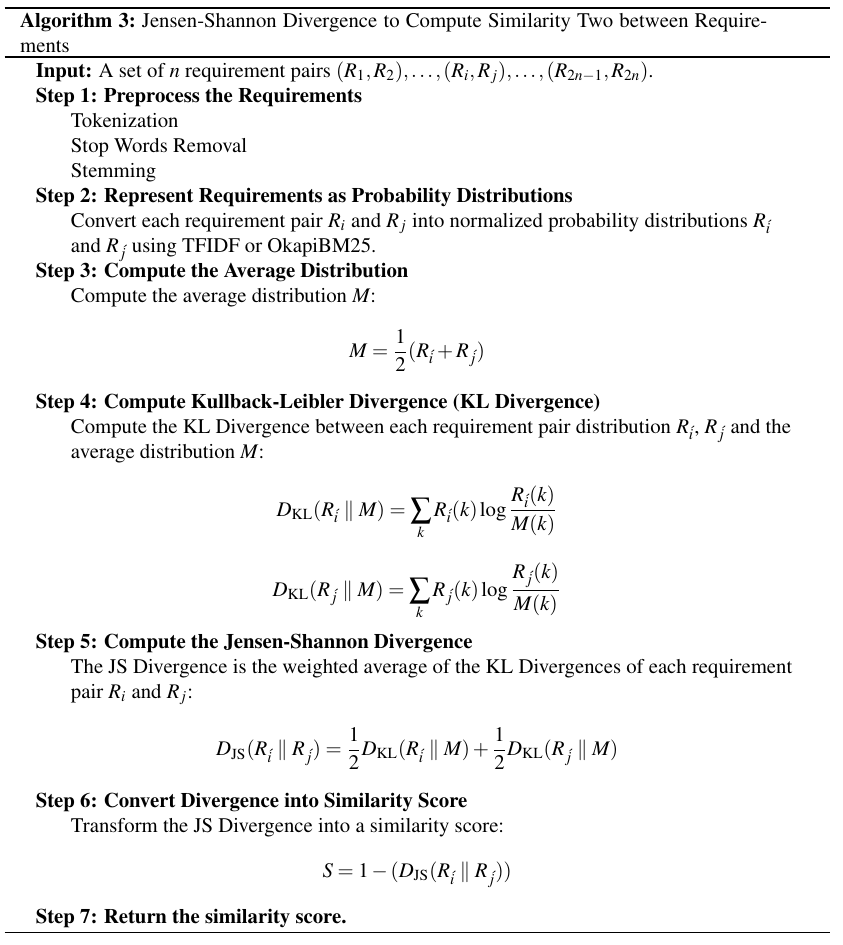} &
        \includegraphics[width=0.45\textwidth]{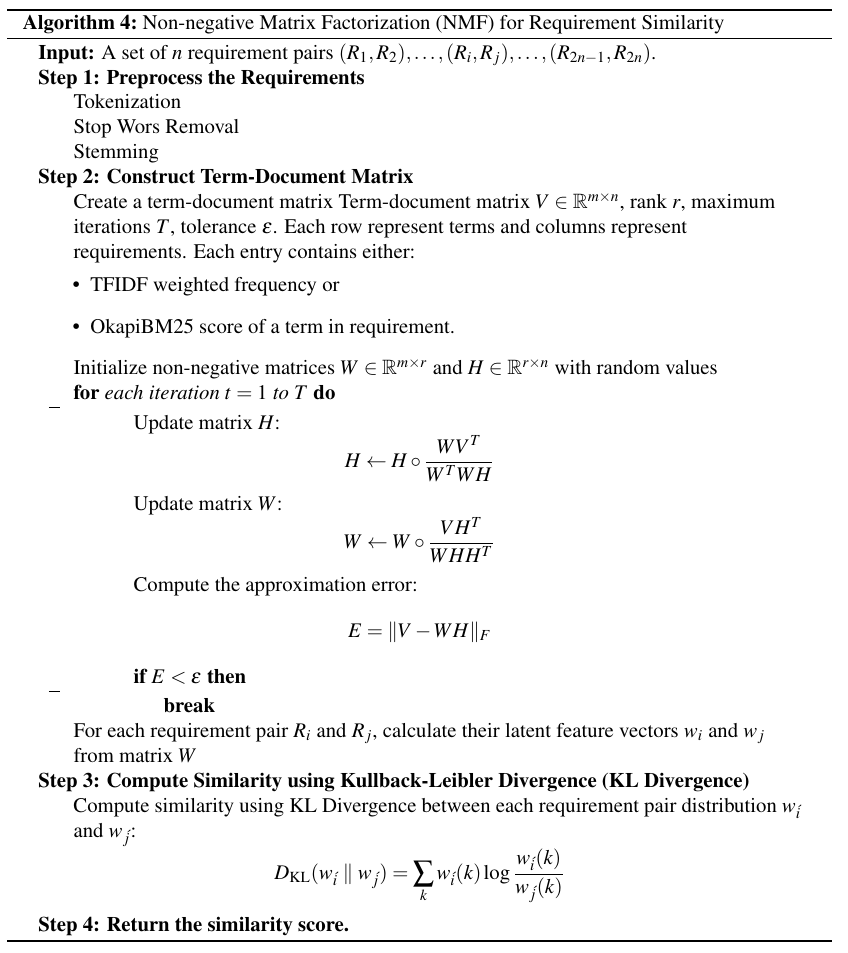} \\
        \textbf{JSD} & \textbf{NMF} \\[1pt]
    \end{tabular}
    \caption{Comparison of different algorithms: VSM, LSI, JSD, and NMF.}
    \label{algos}
\end{figure*}

\begin{figure}
    \centering
    \includegraphics[width=\textwidth]{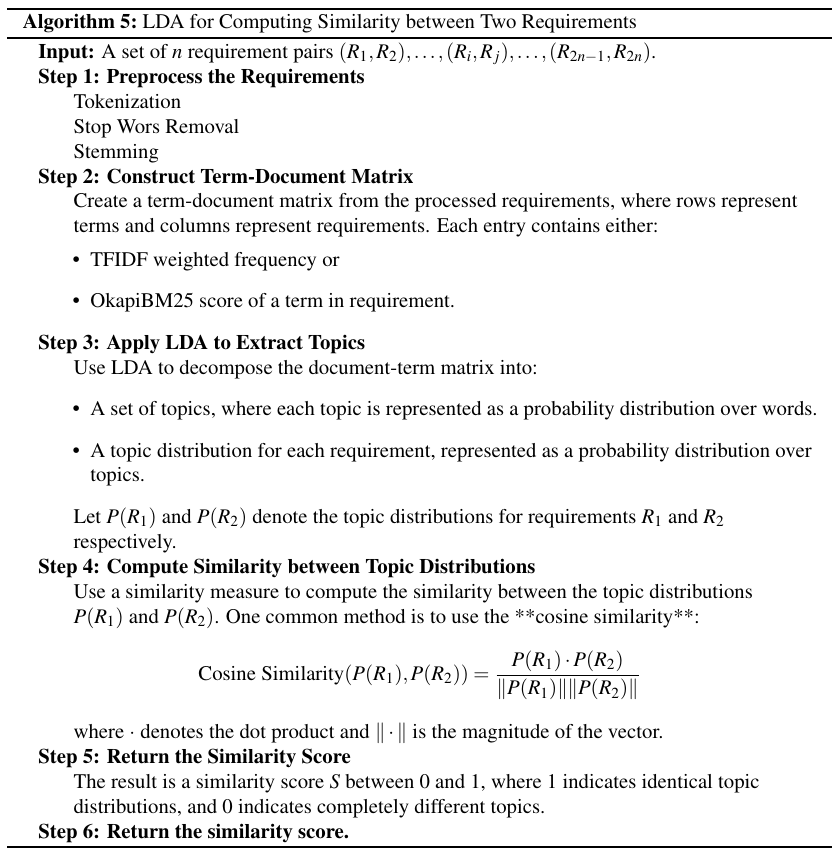}
    \caption{LDA}
    \label{algo5}
\end{figure}

This section offers insights of traditional representation learning and similarity computation based predictive pipelines. These predictive pipelines working paradigm comprises of three distinct stages including statistical representation learning, similarity computation, and classification. First stage involves transformation of requirements text into statistical vectors using TFIDF \citep{saleem7mlr} or OkapiBM25 \citep{saleem7mlr} (in section 1 of supplementary file). The second stage utilizes 10 distinct traditional similarity methods  to compute similarity scores between transformed vectors of requirement pairs. The five main algorithms of similarity computation include Vector Space Model (VSM) \citep{turney2010frequency}, Latent Semantic Indexing (LSI) \citep{kontostathis2006framework}, Jensen-Shannon Divergence (JS) \citep{JensenShannon}, Non-negative Matrix Factorization (NMF) \citep{NMF}, Latent Dirichlet Allocation (LDA) \citep{LDA}. These methods are combined to develop 5 additioal hybird methods namely NMF+LDA, JS+LDA, VSM+NMF, JS+NMF and VSM+JS. Pusudo code of VSM, LSI, JS, NMF and LDA technqiues is shown in Figures  \ref{algos} (a), \ref{algos} (b), \ref{algos} (c), \ref{algos} da) and \ref{algo5}, respectively. Each method offers unique strengths to capture semantic relationships and similarities between requirements. For instance, VSM represents projects requirements in a high-dimensional space, allowing for efficient similarity computation. LSI extends VSM by applying Singular Value Decomposition to reduce dimensionality and capture latent semantic relationships between words. JSI provides a symmetric measure of similarity between probability distributions, while NMF focuses on decomposing non-negative matrices to extract meaningful features. LDA, on the other hand, is particularly useful for uncovering latent entities in requirements.

To describe the concept of similarity computation through traditional methods, lets consider a sample dataset $D = \{(R_i, R_j) \mid i, j \in \{1, 2, \ldots, n\}\}$. In this dataset each sample represents a requirement pair $R_i, R_j$ and each requirement contains a set of words i.e $R_i \in  {W_1,W_2,W_3,……,W_n}$. In the first stage, TFIDF and OkapiBM25 generates statistical vectors of requirements text. Subsequently, theses vectors are passed to 10 distinct similarity methods which produce statistical vector shown in Equation \ref{eq:similarity_matrix}.

\begin{equation}
\label{eq:similarity_matrix}
\begin{aligned}
 & (SR_{1,1}: S_{VSM}^{\text{TFIDF}}, \ldots, S_{VSM+JS}^{\text{TFIDF}}, S_{VSM}^{\text{Okapi}}, \ldots, S_{VSM+JS}^{\text{Okapi}}, \\
 & \hspace{0.2in} \vdots \hspace{0.3in} \vdots \hspace{0.55in} \vdots \hspace{0.4in} \vdots \hspace{0.55in} \vdots \hspace{0.28in}, \\
 & SR_{i,j}: S_{VSM}^{\text{TFIDF}}, \ldots, S_{VSM+JS}^{\text{TFIDF}}, S_{VSM}^{\text{Okapi}}, \ldots, S_{VSM+JS}^{\text{Okapi}}, \\
 & \hspace{0.2in} \vdots \hspace{0.3in} \vdots \hspace{0.55in} \vdots \hspace{0.4in} \vdots \hspace{0.55in} \vdots \hspace{0.28in}, \\
 & SR_{n,n}: S_{VSM}^{\text{TFIDF}}, \ldots, S_{VSM+JS}^{\text{TFIDF}}, S_{VSM}^{\text{Okapi}}, \ldots, S_{VSM+JS}^{\text{Okapi}}) 
\end{aligned}
\end{equation}

 In Equation \ref{eq:similarity_matrix}, $S_{VSM}^{\text{TFIDF}}$ represents the similarity score computed using TFIDF representation and VSM similarity method and subsequent dimensions correspond to other unique combinations of representation and similarity methods.  The generated similarity vectors are passed to 16 different machine learning classifiers namely Gaussian Process (GP) \citep{rasmussen2006gaussian}, Quadratic Discriminant (QD) \citep{friedman1989regularized}, K-Nearest Neighbors (KNN) \citep{cover1967nearest}, Gaussian Naive Bayes (GNB) \citep{john1995estimating}, Bernoulli Naive Bayes (BNB) \citep{john1995estimating}, Support Vector Machines (SVM) \citep{cortes1995support}, Logistic Regression (LogR) \citep{cox1958regression}, Multi-Layer Perceptron (MLP) \citep{rumelhart1986learning}, Decision Trees (DT) \citep{breiman1984classification}, Random Forests (RF) \citep{breiman2001random}, AdaBoost \citep{freund1997decision}, Gradient Boosting (Gboost) \citep{friedman2001greedy}, XGBoost \citep{chen2016xgboost}, CatBoost \citep{prokhorenkova2018catboost}, Histogram-based Gradient Boosting (Histgboost) \citep{guryanov2019histogram}, and LightGBM \citep{ke2017lightgbm}.

To evaluate whether TFIDF representation based computed similarity scores and classifiers outperform those based on OKAPI-BM25 representation, or if a combination of both yields superior results, the framework supports the development of three distinct types of predictive pipelines:
\begin{enumerate}
    \item TFIDF Representation based predictive Pipelines: Utilizes TFIDF representation with 10 similarity methods and 16 classifiers.
    \item OKAPI-BM25 Representation based predictive Pipelines:  Employs OKAPI-BM25 representation with 10 similarity methods and 16 classifiers.
    \item TFIDF and OKAPI-BM25 Representations based predictive Pipelines: Concatenates similarity vectors computed through TFIDF and OKAPI-BM25 representations and passed to 16 distinct classifiers.
\end{enumerate}

This structured approach enables a comprehensive comparison of performance across the three strategies predictive pipelines.

\subsubsection{Large Language Models and Traditional Similarity Techniques based Predictive Pipelines}\label{Similarity-Computation-through-Large-Language- models}
In addition to traditional methods for similarity computation of requirement pairs, PassionNet framework  also offers language model powered similarity computation using 10 different LLMs. These LLMs include: ALBERT \citep{lan2019albert}, BERT \citep{devlin2018bert}, BART \citep{lewis2019bart}, DeBERTa \citep{he2020deberta}, Electra \citep{clark2020electra} GPT \citep{radford2019language}, Longformer \citep{beltagy2020longformer}, RoBERTa \citep{liu2019roberta}, XLM \citep{barbieri2021xlm} and XLNet \citep{yang2019xlnet}. Each requirement pair $R_i$ and $R_j$ is concatenated to form a unified input with a separator token [SEP] inserted between them.
Afterwards, unified input is fed to 10 distinct LLMs and each LLM generates a similarity score $S_{LLM}$ that reflects its unique interpretation of similarity in a requirement pair. These scores collectively make a 10-dimensional similarity vector of requirement pair $R_{ij}$, which can be denoted as $S_{i,j{LLM}}={S_{ALBERT}, S_{BERT}, S_{BART}, S_{DeBERTa},S_{Electra},S_{Longformer},S_{RoBERTa},S_{XLM},S_{XLNet}}$. Here, $S_{roberta}$ represents similarity score computed using pretrained RoBERTa language model.  The generated 10 dimensional similarity vectors are passed to 16 distinct classifiers. Within these predictive pipelines language models extracts text contextual information to compute similarity between requirements pairs and classifiers extracts discriminative and informative patterns from similarity vectors to discriminate requirements pairs into duplicate,  and confilicting, neutral classes. 
In addition, framework leverages traditional representation learning methods and similarity methods based similarity vectors and LLMs similarity vectors combine potential to develop predictive pipelines as follows: 

\begin{enumerate}
    \item LLM similarity based predictive pipelines: Utilized 10-dimensional similarity vector with 16 ML classifiers.
    \item TFIDF + LLMs similarity based predictive pipelines: 16 ML classifiers takes Concatenated similarity vectors computed through LLMS and TFIDF representation with  tradiational similarity computation methods
    \item OkapiBM25 + LLMs similarity based predictive pipelines: 16 ML classifiers takes Concatenated similarity vectors computed through LLMS and OkapiBM25 representation with  tradiational similarity computation methods
    \item Multimodel predictive Pipelines: Similarity vectors concatenation of all three methods (TFIDF + OkapiBM25 + LLMs) and 16 distinct classifiers
\end{enumerate}

A large-scale experimental evaluation of these predictive pipelines offers valuable insights into whether predictive pipelines based solely on similarity vectors derived from Large Language Models (LLMs) deliver superior performance, or combination of traditional similarity vectors and LLM-based vectors yields better results.

\subsection{MultiModel Similarity Knowledge Driven and Language Model Context Aware Predictive Pipelines}

Third type of predictive pipelines utilized diverse types of multimodel similarity knowledge with LLM based context. To extract LLM-based context for requirement pairs, the framework concatenates each pair into a single input with a [SEP] token between requirements. This unified representation enables the LLM to  capture contextual and semantic relations of requirement pairs. Specifically, it extracts [CLS] token representation from 8 different pre-trained LLMs including; ALBERT \citep{lan2019albert}, BERT \citep{devlin2018bert}, DeBERTa \citep{he2020deberta}, Electra \citep{clark2020electra} GPT \citep{radford2019language}, Longformer \citep{beltagy2020longformer}, RoBERTa \citep{liu2019roberta} and XLNet \citep{yang2019xlnet}. The extracted high-dimensional [CLS] token representations is subjected to Principal Component Analysis (PCA) \citep{abdi2010principal} for dimensionality reduction to 5 lower-dimensions (8, 16, 32, 64, and 128). The reduced CLS representations is concatenated with the previously generated multimodel similarity knowledge. For example, a 8-dimensional CLS representation of GPT2 is combined with a 30-dimensional multimodel similarity knowledge vector to produce a 38-dimensional vector for each requirement pair. This process creates enriched similarity vectors that incorporate both LLM-based semantic understanding as well as traditional and advanced similarity assessment which are fed to 16 ML classifiers. By leveraging multiple LLM architectures and varying dimensionality reductions, the framework aims to capture a comprehensive range of semantic and structural similarities between requirement pairs, potentially improving the accuracy and robustness of conflict/duplicate requirement identification.

\section{Benchmark Datasets}
\begin{figure}{htbp}
    \centering
    \includegraphics[width=0.99\textwidth]{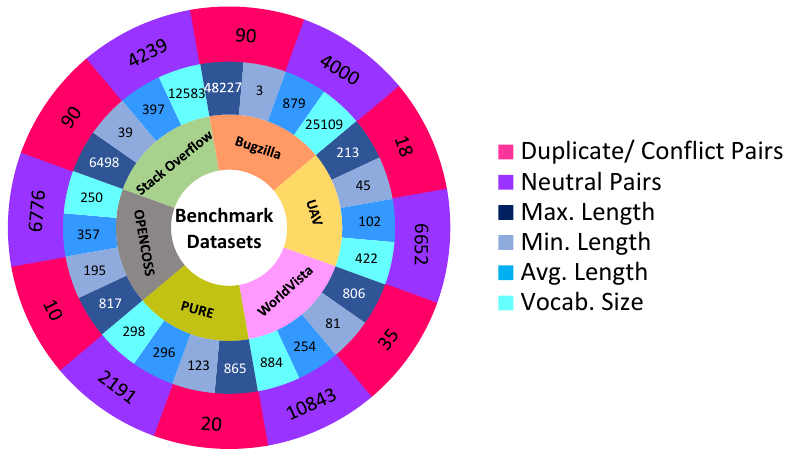}
    \caption{Description of Conflict and Duplicate Detection Datasets}
    \label{dataset-stats}
\end{figure}

To demonstrate proposed PassionNet framework robustness and broad applicability, we conduct a comprehensive evaluation on six publicly available benchmark datasets. Specifically,  2 duplicate requirements datasets (Stack Overflow \citep{helmeczi2023few, malik2023data},  Bugzilla \citep{helmeczi2023few, malik2023data}) and 4 conflicting requirements datasets namely (UAV \citep{malik2023data}, WorldVista \citep{malik2023data}, PURE \citep{malik2023data} and OPENCOSS \citep{malik2023data}). Figure \ref{dataset-stats} briefly describes these datasets statistics in terms of total number of samples, samples distribution across each class, vocabulary size, and maximum, minimum and average lengths of requirements. A brief description of these datasets is provided below.

\begin{itemize}

\item{\textbf{Unmanned Aerial Vehicle (UAV)} 
 The University of Notre Dame released UAV dataset in 2021, which contains requirements related to UAV control system operations \citep{guo2021automatically, cleland2018dronology}. Typically, requirements of UAV control system operations precede main clause, which are represented by verbs such as ``send," ``receive," or ``provide" \citep{guo2021automatically}. In some scenarios, these requirements also contain additional elements, such as ``to do" or ``to make". 
Moreover, requirements in UAV dataset adhere to EARS (Easy Approach to Requirements Syntax) format \citep{mavin2009easy}. This dataset consists of 6,670 requirements pairs labeled as either conflict or neutral. }

\item{\textbf{WorldVista} 
WorldVista \url{https://worldvista.org/Documentation} dataset contains software requirements related to health management system. This system tracks patient information throughout entire process, from hospital admission to discharge. The requirements are written in natural language by incorporating healthcare terminologies and have a simple sentence structure. The WorldVista dataset comprises of 10,878 requirement pairs, each labeled as either conflict or neutral.}

\item{\textbf{PURE} 
In 2023, Malik et al. \citep{malik2023data} developed pure dataset to identify conflict and neutral software requirements pairs by utilizing PURE: publicly available repository of SRS document. This repository contains SRS documents of 79 different projects. The authors only selected two SRS documents namely Thermodynamic System (THEMAS) and web interface for social networks (Mashbot). Raw text of these SRSs is arranged in paragraphs and contains sentences with varying length and structure. To address these issues, authors pre-processed text by breaking down each paragraph into individual sentences. Furthermore, the authors treated each sentence as a single requirement and combined list of requirements into single sentence with comma separated elements. This process ensured that each requirement was represented as a single, clear sentence, with simplifies structure. The dataset consists of 2,211 requirement pairs, with 20 as conflicts and 2,191 as neutral pairs.}

\item{\textbf{OPENCOSS} 
OPENCOSS \url{http://www.opencoss-project.eu} reflects Open Platform for Evolutionary Certification Of Safety-critical Systems for railway, avionics, and automotive industries. Existing studies \citep{malik2023data, malik2023transfer, malik2022supervised} has identified OPENCOSS as a challenging dataset  to discriminate conflict and neutral requirements pairs. This challenge arises due to presence of numerous common words in  both conflict and neutral pairs. This dataset contains 6786 requirement pairs and is highly imbalanced with 10 pairs classified conflict class and 6776 pairs belonging to neutral class.}

\item{\textbf{StackOverflow}
StackOverflow website is dedicated to address software and programming related problems. Due to large number of users, similar questions are frequently posted on website, often focusing on same issue. Despite ongoing efforts to reduce redundant questions that have already been answered, duplicate inquiries still appear on website. To address this challenge, Malik et al. \citep{malik2023data} specifically collected neutral and duplicate questions pairs across various programming languages. To maintain consistency with other datasets, authors considered only 5,000 neutral pairs and 90 duplicate pairs.}


\item{\textbf{Bugzilla} \url{https://bugzilla.mozilla.org/index.cgi} is a widely used, open-source bug tracking system developed by Mozilla Foundation, originally launched in 1998. It is designed to facilitate users in managing and tracking software bugs. However, like other bug tracking systems, Bugzilla also face challenge of duplicate bug reports. Primarily this issue occur due to lack of coordination between users and their unawareness about existing
reports. To facilitate  duplicate detection of bug reports, Malik et al. \citep{malik2023data} scrapped open bug reports from Mozilla’s Bugzilla using the REST API. They categorized 4000 bug reports pairs as neutral and 90 bug reports pairs as duplicate.}
\end{itemize}
\section{Evaluation Measures}
\begin{figure}[htbp]
    \centering
    \includegraphics{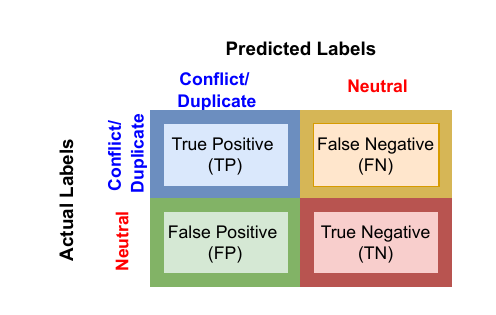}
    \caption{Confusion matrix}
    \label{confusion_matrix}
\end{figure}

To assess performance of proposed framework, we leveraged 6 distinct evaluation measures that are used in existing conflict/duplicate requirements detection predictors \citep{malik2023data, malik2023transfer, malik2022supervised}. Due to highly imbalanced nature of the data, existing predictors have been evaluated over a diverse array of evaluation measures including standard and macro versions of  precision (Pr), recall (R), and F1 score (F1). Primarily standard version of these measures are derived from a confusion matrix, which consists of four components: 1) True Positive (TP), 2) True Negative (TN), 3) False Positive (FP), and 4) False Negative (FN). TP and TN refer to number of correctly predicted conflict/duplicate and neutral requirements pairs, respectively. Contrarily, FP and FN represent number of requirements pairs that are incorrectly classified as conflict/duplicate and neutral pairs, respectively. Figure \ref{confusion_matrix} graphically represents confusion matrix in terms of TP, TN, FP, and FN. Equation \ref{standard_evaluation_measures} illustrates the mathematical expression to compute standard version of precision recall and F1 score evaluation measures.

\begin{equation}\label{standard_evaluation_measures}
f(x)_{standard} = \begin{cases} 
\text{Pr} = \frac{TP}{TP + FP}  \\
\text{R} = \frac{TP}{TP + FN}  \\
\text{F1} = \frac{2 \cdot \text{Pr} \cdot \text{R}}{\text{Pr} + \text{R}} \\

\end{cases}
\end{equation}

Macro variant independently calculates each metrics across all classes and then computes weighted average of matrix with respect to total classes.  Specifically for imbalanced datasets macro variants of metrics address different aspects of class importance. Equation \ref{imbalance} depicts mathematical expression of these metrics in terms of marco variants. 

\begin{equation}\label{imbalance}
f(x)_{\text{imbalance}} = \begin{cases} 
\text{Pr}_{\text{macro}} = \frac{1}{m} \sum_{i=1}^{m} \text{Pr}_i \\
\text{R}_{\text{macro}} = \frac{1}{m} \sum_{i=1}^{m} \text{R}_i  \\
\text{F1}_{\text{macro}} = \frac{2 \cdot \text{Pr}_{\text{macro}} \cdot \text{R}_{\text{macro}}}{\text{Pr}_{\text{macro}} + \text{R}_{\text{macro}}} \\

\end{cases}
\end{equation}

In Equation \ref{imbalance}, $TP_i$, $FP_i$, $TN_i$ and $FN_i$ denote the true positive, false positive, true negative, and false negative counts for class $i$ and $m$ represents total number of classes.

\section{Experimental Setup and Results}
The proposed framework is developed on top of 
8 distinct APIs namely;  scikit-learn \url{https://scikit-learn.org/}, numpy \url{https://numpy.org/}, math \url{https://docs.python.org/3/library/math.html}, scipy \url{https://scipy.org/}, pandas \url{https://pandas.pydata.org/}, matplotlib \url{https://matplotlib.org/}, FastAI \url{https://www.fast.ai/} and pytorch \url{https://pytorch.org/}. Following experimental criteria of existing studies \citep{helmeczi2023few, malik2023data}, we performed experimentation using 3-fold cross-validation setting. In this setting iteratively, one fold is taken as test set and other 2 folds are used for model training. In addition, for each iteration  5 percent of training set is used as validation set for LLM based predictive pipelines. To ensure a fair performance comparison among distinct predictive pipelines, we utilized Optuna \url{https://optuna.org/} to identify the optimal hyperparameters for each pipeline by exploring an extensive hyperparameter search space. In supplementary file, Tables 1 and 2 illustrate ML classifiers and LLMS predictive pipelines hyperparameters and their selected  search spaces. optimize the performance of both ML classifiers and LLMs, we conducted an extensive hyper-parameter search. To obtain representations from LLMs, requirement pairs are  passed to the model separated with a [SEP] token. Afterwards, model's [CLS] token is extracted, which contains semantic relationship between requirements pair. This high-dimensional representation is subsequently reduced to five dimensions using Principal Component Analysis (PCA), creating a compact yet informative embedding that captures the essential features of the requirement pair's relationship. 

The following subsections provides a comprehensive performance analysis of 3 distinct predictive pipelines offered by PassionNet Framework: language model-based, multimodel similarity knowledge driven and a hybrid approach combining LLMs context with multimodel similarity knowledge. Furthermore, it presents a detailed performance comparison of the top-performing predictors of each type of predictive pipelines. Lastly, it provides in-depth performance analysis of existing conflict/duplicate requirements identification  predictors with proposed PassionNet framework.

\subsection{ Performance Benchmarking of Large Language Models based Predictive Pipelines}

This section investigates effectiveness of end-to-end LLM-based predictive pipeline for conflict/duplicate requirements identification. This analysis utilizes 8 diverse LLMs, including ALBERT, BERT, DeBERTa, ELECTRA, GPT, Longformer, RoBERTa and XLNet, across 6 public benchmark datasets. This comprehensive analysis highlight adaptability and strength of different LLMs for conflict/duplicate requirements classification tasks across varying dataset characteristics.

\begin{figure}[htbp]
    \centering
    \includegraphics[width=0.99\textwidth]{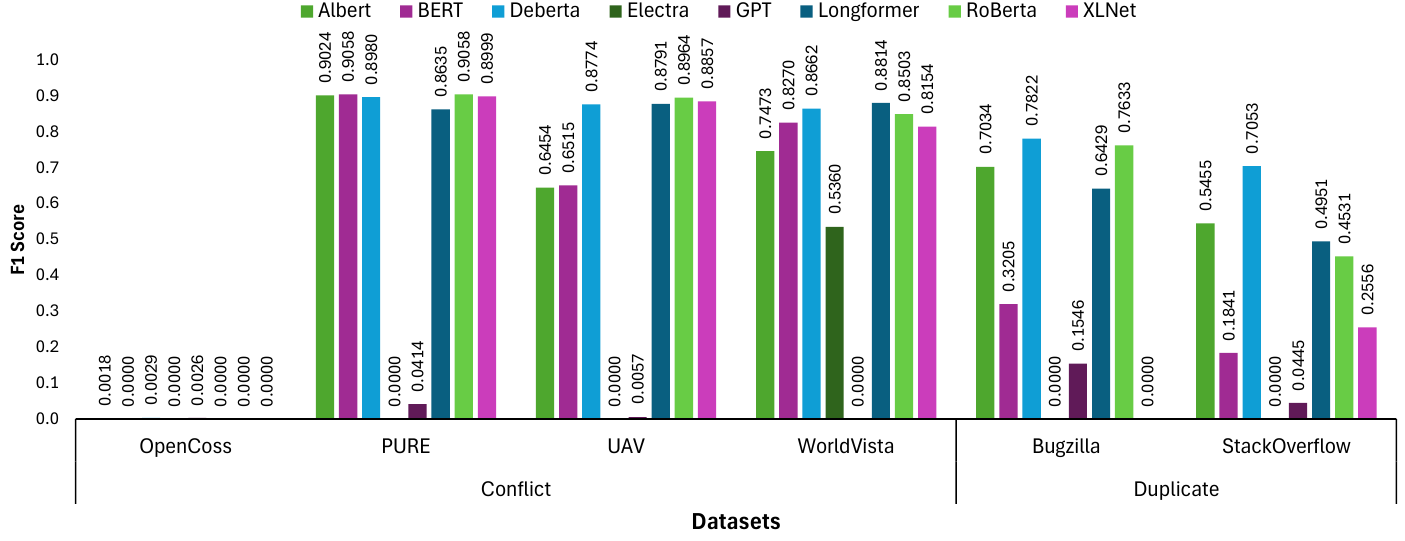}
    \caption{Performance comparison of 8 different LLMs in terms of F1 score of conflict/duplicate class across 6 benchmark datasets of conflict/duplicate requirements identification }
    \label{LLM-benchmarking}
\end{figure}

It is evident from Figure \ref{LLM-benchmarking} over pure dataset most LLMS perform exceptionally well, with five LLMS (Roberta, BERT, ALbert, Xlnet, Deberta) scoring f1 score above 0.89.  On the other hand, 4 LLMs (Roberta, Longformer, DeBERTa and XLNet) excel on UAV dataset by achieving f1 score above 0.87. World dataset demonstrates more consistent performance across majority of LLMs, with 5 LLMs (longformer, deberta, RoBerta, BERT, XLNet) scoring f1 score above 0.80. However,  OPEN dataset stands out as significant challenge, with almost all LLMs scoring very low or nearly zero f1 score that indicates potential difficulties to extracting meaningful patterns from fewer instances of conflict class. Overall, over conflict detection datasets, Roberta remains top-performing over 3 datasets (PURE, UAV, Open) and on WORLD, Roberta falls closely behind the top-performing Longformer with a slight margin. On duplicate detection datasets (BUG, Stack) DeBERTa stands out as top performing predictor. Particularly, on bug dataset Albert roberta performed quite well attaining f1 score above 0.64. However, over stack dataset all LLMs remain fail and secure f1 score below 0.55.

Notably, ELECTRA and GPT failed completely across all  datasets and continually  achieved very low F1 score or zero across conflict/duplicate identification datasets. Both ELECTRA and GPT are sensitive to data imbalance due to their respective training methods and struggle to capture discriminative features of data.  Hence, due to highly imbalanced cases, both models tends to overfit on the neutral class and struggle to accurately identify conflict/duplicate class. DeBERTa and RoBERTa excel in conflict/duplicate detection on imbalanced datasets due to their advanced architectures and fine-tuned training methods. DeBERTa’s disentangled attention mechanism facilitates to capture minute differences in similar text while RoBERTa’s dynamic masking enhances its ability to generalize  semantic discriminations. Hence,  sensitivity to fine-grained semantic and syntactic details along with rich contextual embeddings  of these LLMs prevent overfitting to the majority class.

\subsection{A Comprehensive Performance Analysis of MultiModel Similarity Knowledge Driven Predictive Pipelines}
This section provides comprehensive performance analysis of multimodel similarity knowledge driven predictive pipelines designed  using 10 distinct LLMS, 2 traditional representation learning methods along with 10 traditional similarity methods as described in section \ref{Similarity-Knowledge-informed-Predictive-pipelines}. These predictive pipelines fall into 3 distinct groups: 1) standalone predictors (TFIDF, OkapiBM25, LLM), 2) hybrid predictors ((TFIDF+OkapiBM25, OkapiBM25+LLM, TFIDF+LLM)) 3) multimodel preidtcor (TFIDF+OkapiBM25+LLM). Table \ref{tab:Representation Learning Results} provides performance values of 112 unique multi model similarity knowledge driven predictive pipelines  in terms of conflict/duplicate class f1 score across 6 public benchmark datasets.

It is evident from Table  \ref{tab:Representation Learning Results}  that among standalone predictors, LLM based predictors underperform other standalone traditional predictors ( TFIDF,OkapiBM25). Conversely,  TFIDF-based predictors excel on pure, uav and stackoverflow datasets and also achieved peak performance values of 0.882, 0.854 and 0.557, respectively. In contrast, OkapiBM25-based predictors attain superior performance on  world (0.837), opencoss (0.609) and bug (0.731) datasets. This analysis reveals that standalone traditional predictors (TFIDF, OkapiBM25) are more effective than advanced LLMs to capture similarity patterns between requirements but are sensitive to dataset characteristics.

Among hybrid predictors (TFIDF+OkapiBM25, OkapiBM25+LLM, TFIDF+LLM), combination of traditional methods with LLM enhances its performance which showcases their capability to strengthen advanced techniques. On the other hand, TFIDF+LLM based hybrid predictors outperform OkapiBM25+LLM across all datasets except opencoss and stackoverflow datasets where TFIDF+OkapiBM25 attains higher performance. This indicates that TFIDF provide a more stable foundation for LLM as compared to OkapiBM25. Moreover, hybrid predictors surpasses performance of standalone predictors with slight performance margin except opencoss where OkapiBM25 standouts in terms of f1 score. This suggest that integration of  diverse similarity computation techniques yields more comprehensive spectrum of requirements similarities.
\begin{sidewaystable}
\resizebox{\textwidth}{!}
{
\begin{tabular}{|c|c|c|c|c|c|c|c|c|c|c|c|c|c|c|c|c|c|}
\hline
\textbf{Dataset} & \textbf{\begin{tabular}[c]{@{}c@{}}Representation \\ learning\end{tabular}} & \textbf{GP} & \textbf{QD} & \textbf{KNN} & \textbf{GNB} & \textbf{BNB} & \textbf{SVM} & \textbf{LogR} & \textbf{MLP} & \textbf{DT} & \textbf{RF} & \textbf{adaBoost} & \textbf{Gboost} & \textbf{XGboost} & \textbf{Catboost} & \textbf{Histgboost} & \textbf{Lightgbm} \\ \hline
\multirow{7}{*}{WorldVista} & TFIDF & 0.799 & 0.803 & 0.765 & 0.359 & 0 & 0.805 & 0.79 & 0.798 & 0.793 & 0.823 & 0.79 & 0.69 & 0.772 & 0.755 & 0.773 & 0.767 \\ 
 & OkapiBM25 & 0.824 & 0.815 & 0.81 & 0.421 & 0 & 0.791 & 0.824 & 0.777 & 0.728 & 0.815 & 0.81 & 0.753 & \fbox{0.837} & 0.814 & 0.822 & 0.832 \\ 
 & LLMs & 0.65 & 0.773 & 0.813 & 0.395 & 0 & 0.766 & 0.788 & 0.842 & 0.782 & 0.79 & 0.785 & 0.782 & 0.717 & 0.81 & 0.721 & 0.833 \\ 
 & TFIDF+OkapiBM25 & 0.811 & 0.803 & 0.803 & 0.39 & 0 & 0.824 & 0.79 & 0.839 & 0.81 & 0.824 & 0.81 & 0.758 & 0.828 & 0.817 & 0.839 & 0.814 \\ 
 & TFIDF + LLMs & 0.817 & 0.807 & 0.773 & 0.419 & 0 & 0.779 & 0.79 & 0.805 & 0.823 & 0.779 & 0.808 & 0.774 & 0.838 & 0.836 & 0.761 & \underline{0.845} \\ 
 & OkapiBM25+LLMs & 0.789 & 0.789 & 0.808 & 0.484 & 0 & 0.779 & 0.808 & 0.772 & 0.779 & 0.779 & 0.808 & 0.808 & 0.835 & 0.83 & 0.807 & 0.836 \\ 
 & TFIDF+OkapiBM25+LLMs & 0.798 & 0.819 & 0.803 & 0.431 & 0 & 0.779 & 0.79 & 0.805 & 0.788 & \textbf{0.865} & 0.819 & 0.765 & 0.854 & 0.798 & 0.839 & 0.705 \\ \hline
\multirow{7}{*}{Pure} & TFIDF & 0.801 & 0.88 & \fbox{0.882} & 0.782 & 0 & 0.88 & 0.863 & 0.881 & 0.819 & 0.855 & 0.801 & 0.758 & 0.881 & 0.844 & 0.828 & 0.801 \\ 
 & OkapiBM25 & 0.498 & 0.844 & 0.861 & 0.823 & 0 & 0.88 & 0.857 & 0.885 & 0.8 & 0.817 & 0.817 & 0.781 & 0.781 & 0.817 & 0.817 & 0.817 \\ 
 & LLMs & 0.222 & 0.718 & 0.839 & 0.624 & 0 & 0.834 & 0.665 & 0.844 & 0.635 & 0.795 & 0.689 & 0.617 & 0.795 & 0.733 & 0.775 & 0.792 \\ 
 & TFIDF+OkapiBM25 & 0.837 & 0.855 & 0.881 & 0.808 & 0 & 0.881 & 0.801 & 0.852 & 0.763 & 0.855 & 0.781 & 0.781 & 0.8 & 0.817 & 0.861 & 0 \\ 
 & TFIDF + LLMs & 0.834 & 0.801 & 0.869 & 0.822 & 0 & 0.869 & \underline{0.889} & \underline{0.889} & 0.819 & 0.855 & 0.798 & 0.784 & 0.885 & 0.791 & 0.839 & 0.801 \\ 
 & OkapiBM25+LLMs & 0.814 & 0.857 & 0.872 & 0.837 & 0 & 0.836 & 0.82 & 0.851 & 0.8 & 0.8 & 0.781 & 0.783 & 0.739 & 0.792 & 0.837 & 0.796 \\ 
 & TFIDF+OkapiBM25+LLMs & 0.855 & 0.671 & \textbf{0.906}  & 0.852 & 0 & 0.861 & 0.885 & 0.851 & 0.763 & 0.855 & 0.763 & 0.808 & 0.8 & 0.763 & 0.855 & 0.836 \\ \hline
\multirow{7}{*}{UAV} & TFIDF & 0.792 & 0.846 & 0.853 & 0.402 & 0 & \fbox{0.854} & 0.853 & 0.823 & 0.832 & 0.847 & 0.814 & 0.853 & 0.766 & 0.862 & 0.847 & 0.812 \\ 
 & OkapiBM25 & 0.095 & 0.775 & 0.782 & 0.567 & 0.095 & 0.656 & 0.803 & 0.846 & 0.727 & 0.788 & 0.788 & 0.574 & 0.707 & 0.663 & 0.812 & 0.684 \\ 
 & LLMs & 0 & 0.768 & 0.828 & 0.74 & 0 & 0.791 & 0.776 & 0.822 & 0.767 & 0.763 & 0.841 & 0.756 & 0.777 & 0.81 & 0.761 & 0.747 \\ 
 & TFIDF+OkapiBM25 & 0.792 & 0.823 & 0.823 & 0.481 & 0.095 & 0.823 & 0.851 & 0.852 & 0.832 & 0.847 & 0.768 & 0.77 & 0.726 & 0.768 & 0.798 & 0.847 \\ 
 & TFIDF + LLMs & 0.847 & 0.848 & 0.823 & 0.679 & 0 & 0.833 & 0.864 & 0.836 & 0.801 & 0.767 & \underline{0.859} & 0.803 & 0.803 & 0.792 & 0.763 & 0.737 \\ 
 & OkapiBM25+LLMs & 0.584 & 0.833 & 0.812 & 0.723 & 0.095 & 0.812 & 0.803 & 0.717 & 0.819 & 0.768 & 0.716 & 0.725 & 0.794 & 0.782 & 0.778 & 0.626 \\ 
 & TFIDF+OkapiBM25+LLMs & 0.847 & 0.484 & 0.812 & 0.62 & 0.083 & 0.833 & \textbf{0.873} & 0.701 & 0.867 & 0.803 & 0.716 & 0.734 & 0.803 & 0.803 & 0.687 & 0.752 \\ \hline
\multirow{7}{*}{OpenCoss} & TFIDF & 0 & 0 & 0.222 & 0.093 & 0 & 0.19 & 0 & 0 & 0.489 & 0.415 & 0.489 & 0.413 & 0.413 & 0 & 0 & 0 \\ 
 & OkapiBM25 & 0 & 0 & 0.489 & 0.177 & 0 & 0.19 & 0.489 & 0.433 & \fbox{0.609} & 0.584 & 0.317 & 0.578 & 0 & 0.19 & 0 & 0 \\ 
 & LLMs & 0 & 0 & 0 & 0.126 & 0 & 0 & 0 & 0 & 0.19 & 0 & 0 & 0 & 0.074 & 0.083 & 0 & 0 \\ 
 & TFIDF+OkapiBM25 & 0 & 0 & 0.222 & 0.141 & 0 & 0.333 & 0.489 & 0.148 & 0.508 & 0.489 & \underline{0.578} & 0.486 & 0.389 & 0.19 & 0 & 0 \\ 
 & TFIDF + LLMs & 0 & 0 & 0.457 & 0.125 & 0 & 0 & 0.133 & 0.133 & 0.413 & 0.389 & 0.489 & 0 & 0.167 & 0.167 & 0 & 0 \\ 
 & OkapiBM25+LLMs & 0 & 0 & 0.521 & 0.194 & 0 & 0.468 & 0.389 & 0.562 & 0.546 & 0.25 & 0.333 & 0.3 & 0.182 & 0 & 0 & 0 \\ 
 & TFIDF+OkapiBM25+LLMs & 0 & 0 & 0.222 & 0.154 & 0 & \textbf{0.612} & 0 & 0.522 & 0.444 & 0.222 & 0.489 & 0.367 & 0 & 0 & 0 & 0 \\ \hline
\multirow{7}{*}{Stackoverflow} & TFIDF & 0.484 & 0.513 & \fbox{0.557}  & 0.469 & 0 & 0.493 & 0.552 & 0.559 & 0.466 & 0.518 & 0.489 & 0.542 & 0.524 & 0.515 & 0.532 & 0.537 \\ 
 & OkapiBM25 & 0.368 & 0.495 & 0.482 & 0.437 & 0 & 0.456 & 0.442 & 0.507 & 0.523 & 0.4 & 0.486 & 0.422 & 0.498 & 0.487 & 0.454 & 0.297 \\ 
 & LLMs & 0 & 0 & 0.075 & 0.218 & 0 & 0.079 & 0.042 & 0.16 & 0.17 & 0.104 & 0.063 & 0.076 & 0.152 & 0.021 & 0.08 & 0.114 \\ 
 & TFIDF+OkapiBM25 & 0.471 & 0.505 & \underline{0.568}& 0.488 & 0 & 0.523 & 0.51 & \underline{0.568} & 0.466 & 0.521 & 0.564 & 0.541 & 0.507 & 0.538 & 0.478 & 0.443 \\ 
 & TFIDF + LLMs & 0.486 & 0.521 & 0.482 & 0.471 & 0 & 0.499 & 0.505 & 0.117 & 0.466 & 0.514 & 0.564 & 0.466 & 0.446 & 0.507 & 0.467 & 0.521 \\ 
 & OkapiBM25+LLMs & 0.358 & 0.42 & 0.319 & 0.443 & 0 & 0.467 & 0.547 & 0.551 & 0.522 & 0.427 & 0.472 & 0.5 & 0.464 & 0.5 & 0.39 & 0.43 \\ 
 & TFIDF+OkapiBM25+LLMs & 0.456 & 0.499 & 0.535 & 0.485 & 0 & 0.565 & 0.464 & \textbf{0.577} & 0.466 & 0.515 & 0.564 & 0.555 & 0.469 & 0.521 & 0.464 & 0.323 \\ \hline
\multirow{7}{*}{Bugzilla} & TFIDF & 0.564 & 0.563 & 0.71 & 0.451 & 0 & 0.718 & 0.648 & 0.725 & 0.65 & 0.684 & 0.62 & 0.69 & 0.669 & 0.646 & 0.647 & 0.675 \\ 
 & OkapiBM25 & 0.25 & 0.684 & 0.636 & 0.236 & 0 & 0.705 & 0.637 & 0.704 & 0.661 & 0.689 & 0.69 & 0.701 & 0.704 & 0.727 & 0.684 & \fbox{0.731} \\ 
 & LLMs & 0 & 0 & 0.058 & 0.289 & 0 & 0 & 0.042 & 0.135 & 0.104 & 0.082 & 0.079 & 0 & 0 & 0 & 0 & 0 \\ 
 & TFIDF+OkapiBM25 & 0.584 & 0.545 & 0.684 & 0.36 & 0 & 0.701 & 0.644 & 0.731 & 0.708 & 0.708 & 0.643 & 0.69 & 0.732 & 0.708 & 0.721 & 0.722 \\ 
 & TFIDF + LLMs & 0.568 & 0.542 & 0.674 & 0.432 & 0 & \underline{0.742} & 0.67 & 0.74 & 0.642 & 0.694 & 0.629 & 0.706 & 0.719 & 0.696 & 0.637 & 0.307 \\ 
 & OkapiBM25+LLMs & 0.265 & 0.608 & 0.581 & 0.219 & 0 & 0.732 & 0.649 & 0.712 & 0.666 & 0.662 & 0.657 & 0.721 & 0.699 & 0.72 & 0.705 & 0.728 \\ 
 & TFIDF+OkapiBM25+LLMs & 0.539 & 0.565 & 0.68 & 0.36 & 0 & 0.738 & 0.655 & \textbf{0.764} & 0.685 & 0.7 & 0.629 & 0.719 & 0.744 & 0.696 & 0.711 & 0.737 \\ \hline
\end{tabular}
}
\caption{ Performance analysis of multimodel similarity
knowledge driven predictive pipelines across 6 different datasets}
\label{tab:Representation Learning Results}
\end{sidewaystable}
A through performance analysis of standalone, hybrid and multimodel predictors reveals that mutltimodel predictors (TFIDF+OkapiBM25+LLMs) outperform other predictors on all datasets. This suggests that integrating different similarity approaches captures a more comprehensive requirements relations, leading to better conflict/duplicate identification. Furthermore, the effectiveness of classifiers is influenced by both predictor's nature and dataset's  variability. For instance, SVM, RF and MLP consistently achieve good performance, while classifiers like GNB and BNB often lag behind, with BNB often yielding zero f1scores.   
\subsection{A Comprehensive Performance Analysis of MultiModel Similarity Knowledge Driven and Language Model Context Aware Predictive Pipelines}
This section presents a comprehensive analysis of 640 unique multimodel similarity knowledge driven and language model context aware predictive pipelines. These pipelines utilize CLS token of eight distinct language models  five dimensions (8, 16, 32, 64, or 128) concatenated with a 30dimensional multimodel similarity vector (TFIDF+OkapiBM25+LLM) and fed to 16 ML classifiers.  The performance metrics for these 640 pipelines are detailed in the supplementary file. Table \ref{tab:LLM+similarity-vctor-results} showcases the performance values of the most effective combinations, highlighting the ML classifier and language model CLS token with the optimal reduced dimension for each dataset (World, Pure, UAV, and Open). This approach allows for a nuanced comparison of how different language models and classifiers interact to produce varying levels of predictive accuracy across different datasets.
\newline
For conflict detection task, Longformer+SV achieves the highest overall performance, with a peak accuracy using the MLP classifier across all datasets. This suggests that Longformer's architecture, combined with the additional vector, is particularly effective for conflict identification. 
BERT, Roberta, and Electra also show strong performance in various scenarios. Interestingly, the performance of models varies significantly across different classifiers, indicating that the choice of classifier is crucial for optimal results. Among the classifiers, MLP consistently achieves high accuracy across different models and datasets. SVM  and LR also perform well in many cases. Tree based models like RF and gradient boosting variants (XGBoost, CatBoost) show competitive performance, especially in the World and Pure datasets. The effectiveness of classifiers varies depending on the dataset and model combination, highlighting the importance of careful classifier selection for each specific task.
\begin{sidewaystable}
\resizebox{\textwidth}{!}
{
\begin{tabular}{|l|l|l|llllllllllllllll|}
\hline
\multirow{2}{*}{Dataset} & \multirow{2}{*}{\begin{tabular}[c]{@{}l@{}}LLM CLS token\\ + 30 dimensional vector\end{tabular}} & \multirow{2}{*}{LLMs CLS Token Dimension}                                    & \multicolumn{16}{c|}{Classifier}                                                                                                                                                                                                                                                                                                                                                                                                                                                                                                    \\ 
                         &                                                                                                  &                                                                              & \multicolumn{1}{l|}{GP}    & \multicolumn{1}{l|}{QC}    & \multicolumn{1}{l|}{kNN}         & \multicolumn{1}{l|}{GNB}   & \multicolumn{1}{l|}{BNB}   & \multicolumn{1}{l|}{SVM}         & \multicolumn{1}{l|}{LR}          & \multicolumn{1}{l|}{mlp}                  & \multicolumn{1}{l|}{DT}          & \multicolumn{1}{l|}{RF}          & \multicolumn{1}{l|}{ADB}   & \multicolumn{1}{l|}{gboost}      & \multicolumn{1}{l|}{xgboost}     & \multicolumn{1}{l|}{catboost}    & \multicolumn{1}{l|}{histgboost}  & lightgbm    \\ \hline
\multirow{8}{*}{World}   & Albert +SV                                                                                       & {[}128, 8, 8, 128, 128, 64, 16, 16, 128, 64, 32, 64, 16, 128, 128, 128{]}    & \multicolumn{1}{l|}{0.541} & \multicolumn{1}{l|}{0.763} & \multicolumn{1}{l|}{0.798}       & \multicolumn{1}{l|}{0.482} & \multicolumn{1}{l|}{0.681} & \multicolumn{1}{l|}{0.850}       & \multicolumn{1}{l|}{0.850}       & \multicolumn{1}{l|}{0.846}                & \multicolumn{1}{l|}{0.833}       & \multicolumn{1}{l|}{0.881}       & \multicolumn{1}{l|}{0.808} & \multicolumn{1}{l|}{0.839}       & \multicolumn{1}{l|}{0.856}       & \multicolumn{1}{l|}{0.861}       & \multicolumn{1}{l|}{0.891} & 0.869       \\ 
                         & BERT-large+ SV                                                                                   & {[}64, 8, 32, 128, 128, 64, 32, 128, 64, 128, 128, 128, 128, 64, 128, 128{]} & \multicolumn{1}{l|}{0.727} & \multicolumn{1}{l|}{0.813} & \multicolumn{1}{l|}{0.807}       & \multicolumn{1}{l|}{0.474} & \multicolumn{1}{l|}{0.712} & \multicolumn{1}{l|}{0.909} & \multicolumn{1}{l|}{0.834}       & \multicolumn{1}{l|}{0.899}                & \multicolumn{1}{l|}{0.840}       & \multicolumn{1}{l|}{0.879}       & \multicolumn{1}{l|}{0.889} & \multicolumn{1}{l|}{0.846}       & \multicolumn{1}{l|}{0.862}       & \multicolumn{1}{l|}{0.882}       & \multicolumn{1}{l|}{0.925}& 0.893       \\ 
                         & Debrta+SV                                                                                        & {[}8, 16, 32, 128, 128, 64, 8, 64, 16, 128, 64, 64, 8, 64, 64, 64{]}         & \multicolumn{1}{l|}{0.817} & \multicolumn{1}{l|}{0.824} & \multicolumn{1}{l|}{0.878} & \multicolumn{1}{l|}{0.514} & \multicolumn{1}{l|}{0.775} & \multicolumn{1}{l|}{0.869}       & \multicolumn{1}{l|}{0.790}       & \multicolumn{1}{l|}{0.861}                & \multicolumn{1}{l|}{0.796}       & \multicolumn{1}{l|}{0.881}       & \multicolumn{1}{l|}{0.831} & \multicolumn{1}{l|}{0.819}       & \multicolumn{1}{l|}{0.846}       & \multicolumn{1}{l|}{0.893}       & \multicolumn{1}{l|}{0.893}       & 0.874       \\
                         & Electra+SV                                                                                       & {[}8, 8, 128, 128, 128, 16, 8, 16, 8, 8, 32, 16, 8, 128, 32, 16{]}           & \multicolumn{1}{l|}{0.830} & \multicolumn{1}{l|}{0.777} & \multicolumn{1}{l|}{0.855}       & \multicolumn{1}{l|}{0.476} & \multicolumn{1}{l|}{0.728} & \multicolumn{1}{l|}{0.817}       & \multicolumn{1}{l|}{0.790}       & \multicolumn{1}{l|}{0.857}                & \multicolumn{1}{l|}{0.806}       & \multicolumn{1}{l|}{0.892}       & \multicolumn{1}{l|}{0.863} & \multicolumn{1}{l|}{0.869}       & \multicolumn{1}{l|}{0.902}       & \multicolumn{1}{l|}{0.909} & \multicolumn{1}{l|}{0.909} & 0.895       \\ 
                         & GPT+ SV                                                                                          & {[}8, 8, 8, 8, 64, 8, 8, 16, 8, 64, 16, 16, 64, 128, 8, 64{]}                & \multicolumn{1}{l|}{0.781} & \multicolumn{1}{l|}{0.835} & \multicolumn{1}{l|}{0.785}       & \multicolumn{1}{l|}{0.371} & \multicolumn{1}{l|}{0.058} & \multicolumn{1}{l|}{0.824}       & \multicolumn{1}{l|}{0.779}       & \multicolumn{1}{l|}{0.788}                & \multicolumn{1}{l|}{0.823}       & \multicolumn{1}{l|}{0.874}       & \multicolumn{1}{l|}{0.815} & \multicolumn{1}{l|}{0.825}       & \multicolumn{1}{l|}{0.885} & \multicolumn{1}{l|}{0.829}       & \multicolumn{1}{l|}{0.873}       & 0.719       \\ 
                         & Longformer+SV                                                                                    & {[}16, 8, 64, 128, 128, 128, 8, 128, 32, 8, 8, 16, 64, 16, 8, 64{]}          & \multicolumn{1}{l|}{0.830} & \multicolumn{1}{l|}{0.861} & \multicolumn{1}{l|}{0.910}       & \multicolumn{1}{l|}{0.476} & \multicolumn{1}{l|}{0.617} & \multicolumn{1}{l|}{0.891}       & \multicolumn{1}{l|}{0.790}       & \multicolumn{1}{l|}{\textbf{0.939}} & \multicolumn{1}{l|}{0.848}       & \multicolumn{1}{l|}{0.879}       & \multicolumn{1}{l|}{0.852} & \multicolumn{1}{l|}{0.889}       & \multicolumn{1}{l|}{0.883}       & \multicolumn{1}{l|}{0.879}       & \multicolumn{1}{l|}{0.908}       & 0.894       \\
                         & Roberta-base+SV                                                                                  & {[}8, 8, 32, 128, 128, 8, 8, 64, 64, 8, 32, 16, 32, 16, 128, 32{]}           & \multicolumn{1}{l|}{0.830} & \multicolumn{1}{l|}{0.866} & \multicolumn{1}{l|}{0.865}       & \multicolumn{1}{l|}{0.470} & \multicolumn{1}{l|}{0.605} & \multicolumn{1}{l|}{0.898} & \multicolumn{1}{l|}{0.790}       & \multicolumn{1}{l|}{0.852}                & \multicolumn{1}{l|}{0.832}       & \multicolumn{1}{l|}{0.874}       & \multicolumn{1}{l|}{0.881} & \multicolumn{1}{l|}{0.898} & \multicolumn{1}{l|}{0.866}       & \multicolumn{1}{l|}{0.876}       & \multicolumn{1}{l|}{0.885}       & 0.865       \\
                         & Xlnet+ SV                                                                                        & {[}128, 8, 128, 128, 128, 8, 32, 8, 128, 64, 16, 8, 16, 16, 128, 64{]}       & \multicolumn{1}{l|}{0.460} & \multicolumn{1}{l|}{0.835} & \multicolumn{1}{l|}{0.650}       & \multicolumn{1}{l|}{0.449} & \multicolumn{1}{l|}{0.563} & \multicolumn{1}{l|}{0.806}       & \multicolumn{1}{l|}{0.823}       & \multicolumn{1}{l|}{0.842}                & \multicolumn{1}{l|}{0.801}       & \multicolumn{1}{l|}{0.879}       & \multicolumn{1}{l|}{0.841} & \multicolumn{1}{l|}{0.829}       & \multicolumn{1}{l|}{0.887}       & \multicolumn{1}{l|}{0.864}       & \multicolumn{1}{l|}{0.896} & 0.864       \\ \hline
\multirow{8}{*}{Pure}    & Albert +SV                                                                                       & {[}64, 128, 16, 64, 128, 8, 64, 128, 16, 32, 64, 16, 8, 16, 128, 8{]}        & \multicolumn{1}{l|}{0.766} & \multicolumn{1}{l|}{0.547} & \multicolumn{1}{l|}{0.872}       & \multicolumn{1}{l|}{0.852} & \multicolumn{1}{l|}{0.777} & \multicolumn{1}{l|}{0.905}       & \multicolumn{1}{l|}{0.918} & \multicolumn{1}{l|}{0.947}                & \multicolumn{1}{l|}{0.763}       & \multicolumn{1}{l|}{0.872}       & \multicolumn{1}{l|}{0.800} & \multicolumn{1}{l|}{0.811}       & \multicolumn{1}{l|}{0.872}       & \multicolumn{1}{l|}{0.817}       & \multicolumn{1}{l|}{0.892}       & 0.836       \\  
                         & BERT-large+ SV                                                                                   & {[}16, 32, 16, 16, 128, 32, 128, 32, 32, 8, 32, 16, 16, 32, 8, 8{]}          & \multicolumn{1}{l|}{0.377} & \multicolumn{1}{l|}{0.766} & \multicolumn{1}{l|}{0.867}       & \multicolumn{1}{l|}{0.852} & \multicolumn{1}{l|}{0.728} & \multicolumn{1}{l|}{0.900} & \multicolumn{1}{l|}{0.844}       & \multicolumn{1}{l|}{0.863}                & \multicolumn{1}{l|}{0.811}       & \multicolumn{1}{l|}{0.872}       & \multicolumn{1}{l|}{0.857} & \multicolumn{1}{l|}{0.825}       & \multicolumn{1}{l|}{0.872}       & \multicolumn{1}{l|}{0.836}       & \multicolumn{1}{l|}{0.861}       & 0.872       \\ 
                         & Debrta+SV                                                                                        & {[}8, 8, 32, 8, 64, 128, 16, 32, 8, 16, 128, 8, 16, 64, 64, 64{]}            & \multicolumn{1}{l|}{0.820} & \multicolumn{1}{l|}{0.592} & \multicolumn{1}{l|}{0.872}       & \multicolumn{1}{l|}{0.852} & \multicolumn{1}{l|}{0.863} & \multicolumn{1}{l|}{0.928}       & \multicolumn{1}{l|}{0.928}       & \multicolumn{1}{l|}{0.933}          & \multicolumn{1}{l|}{0.781}       & \multicolumn{1}{l|}{0.872}       & \multicolumn{1}{l|}{0.837} & \multicolumn{1}{l|}{0.833}       & \multicolumn{1}{l|}{0.872}       & \multicolumn{1}{l|}{0.872}       & \multicolumn{1}{l|}{0.841}       & 0.857       \\ 
                         & Electra+SV                                                                                       & {[}8, 8, 32, 8, 64, 8, 16, 64, 64, 8, 32, 32, 8, 8, 8, 16{]}                 & \multicolumn{1}{l|}{0.837} & \multicolumn{1}{l|}{0.550} & \multicolumn{1}{l|}{ 0.928} & \multicolumn{1}{l|}{0.852} & \multicolumn{1}{l|}{0.506} & \multicolumn{1}{l|}{0.869}       & \multicolumn{1}{l|}{0.857}       & \multicolumn{1}{l|}{0.906}                & \multicolumn{1}{l|}{0.769}       & \multicolumn{1}{l|}{0.872}       & \multicolumn{1}{l|}{0.837} & \multicolumn{1}{l|}{0.792}       & \multicolumn{1}{l|}{0.837}       & \multicolumn{1}{l|}{0.837}       & \multicolumn{1}{l|}{0.851}       & 0.836       \\  
                         & GPT+ SV                                                                                          & {[}8, 8, 8, 8, 128, 8, 8, 128, 8, 8, 32, 32, 8, 16, 64, 8{]}                 & \multicolumn{1}{l|}{0.703} & \multicolumn{1}{l|}{0.550} & \multicolumn{1}{l|}{0.731}       & \multicolumn{1}{l|}{0.837} & \multicolumn{1}{l|}{0.029} & \multicolumn{1}{l|}{0.911} & \multicolumn{1}{l|}{0.889}       & \multicolumn{1}{l|}{0.909}                & \multicolumn{1}{l|}{0.827}       & \multicolumn{1}{l|}{0.872}       & \multicolumn{1}{l|}{0.763} & \multicolumn{1}{l|}{0.827}       & \multicolumn{1}{l|}{0.902}       & \multicolumn{1}{l|}{0.819}       & \multicolumn{1}{l|}{0.783}       & 0.872       \\ 
                         & Longformer+SV                                                                                    & {[}8, 8, 8, 8, 128, 32, 64, 32, 8, 8, 16, 128, 64, 64, 128, 128{]}           & \multicolumn{1}{l|}{0.872} & \multicolumn{1}{l|}{0.550} & \multicolumn{1}{l|}{0.869}       & \multicolumn{1}{l|}{0.852} & \multicolumn{1}{l|}{0.538} & \multicolumn{1}{l|}{0.880}       & \multicolumn{1}{l|}{0.903}       & \multicolumn{1}{l|}{\textbf{0.974}} & \multicolumn{1}{l|}{0.763}       & \multicolumn{1}{l|}{0.872}       & \multicolumn{1}{l|}{0.756} & \multicolumn{1}{l|}{0.781}       & \multicolumn{1}{l|}{0.872}       & \multicolumn{1}{l|}{0.826}       & \multicolumn{1}{l|}{0.922}       & 0.902       \\  
                         & Roberta-base+SV                                                                                  & {[}8, 8, 16, 8, 128, 8, 16, 128, 8, 8, 16, 8, 8, 8, 32, 8{]}                 & \multicolumn{1}{l|}{0.872} & \multicolumn{1}{l|}{0.550} & \multicolumn{1}{l|}{0.906}       & \multicolumn{1}{l|}{0.852} & \multicolumn{1}{l|}{0.543} & \multicolumn{1}{l|}{0.869}       & \multicolumn{1}{l|}{0.947} & \multicolumn{1}{l|}{0.947}          & \multicolumn{1}{l|}{0.763}       & \multicolumn{1}{l|}{0.872}       & \multicolumn{1}{l|}{0.837} & \multicolumn{1}{l|}{0.815}       & \multicolumn{1}{l|}{0.872}       & \multicolumn{1}{l|}{0.811}       & \multicolumn{1}{l|}{0.851}       & 0.872       \\
                         & Xlnet+ SV                                                                                        & {[}32, 8, 32, 32, 32, 64, 8, 32, 32, 128, 64, 64, 16, 128, 16, 8{]}          & \multicolumn{1}{l|}{0.733} & \multicolumn{1}{l|}{0.536} & \multicolumn{1}{l|}{0.823}       & \multicolumn{1}{l|}{0.852} & \multicolumn{1}{l|}{0.713} & \multicolumn{1}{l|}{0.914}       & \multicolumn{1}{l|}{0.947} & \multicolumn{1}{l|}{0.889}                & \multicolumn{1}{l|}{0.781}       & \multicolumn{1}{l|}{0.902}       & \multicolumn{1}{l|}{0.837} & \multicolumn{1}{l|}{0.830}       & \multicolumn{1}{l|}{0.928}       & \multicolumn{1}{l|}{0.881}       & \multicolumn{1}{l|}{0.922}       & 0.847       \\ \hline
\multirow{8}{*}{UAV}     & Albert +SV                                                                                       & {[}128, 8, 64, 128, 128, 32, 128, 128, 8, 8, 128, 128, 16, 8, 16, 32{]}      & \multicolumn{1}{l|}{0.767} & \multicolumn{1}{l|}{0.262} & \multicolumn{1}{l|}{0.949}       & \multicolumn{1}{l|}{0.743} & \multicolumn{1}{l|}{0.874} & \multicolumn{1}{l|}{0.941} & \multicolumn{1}{l|}{0.918}       & \multicolumn{1}{l|}{0.919}                & \multicolumn{1}{l|}{0.847}       & \multicolumn{1}{l|}{0.833}       & \multicolumn{1}{l|}{0.847} & \multicolumn{1}{l|}{0.914}       & \multicolumn{1}{l|}{0.828}       & \multicolumn{1}{l|}{0.879}       & \multicolumn{1}{l|}{0.819}       & 0.884       \\ 
                         & BERT-large+ SV                                                                                   & {[}128, 8, 16, 128, 64, 8, 16, 8, 16, 32, 16, 64, 64, 8, 64, 64{]}           & \multicolumn{1}{l|}{0.312} & \multicolumn{1}{l|}{0.262} & \multicolumn{1}{l|}{0.652}       & \multicolumn{1}{l|}{0.630} & \multicolumn{1}{l|}{0.501} & \multicolumn{1}{l|}{0.788}       & \multicolumn{1}{l|}{0.859}       & \multicolumn{1}{l|}{0.827}                & \multicolumn{1}{l|}{0.884}       & \multicolumn{1}{l|}{0.828}       & \multicolumn{1}{l|}{0.878} & \multicolumn{1}{l|}{0.852}       & \multicolumn{1}{l|}{0.862}       & \multicolumn{1}{l|}{0.909} & \multicolumn{1}{l|}{0.836}       & 0.859       \\ 
                         & Debrta+SV                                                                                        & {[}8, 8, 8, 128, 128, 64, 64, 32, 32, 8, 8, 16, 8, 16, 8, 128{]}             & \multicolumn{1}{l|}{0.823} & \multicolumn{1}{l|}{0.262} & \multicolumn{1}{l|}{0.859}       & \multicolumn{1}{l|}{0.707} & \multicolumn{1}{l|}{0.587} & \multicolumn{1}{l|}{0.801}       & \multicolumn{1}{l|}{0.836}       & \multicolumn{1}{l|}{0.909}         & \multicolumn{1}{l|}{0.862}       & \multicolumn{1}{l|}{0.828}       & \multicolumn{1}{l|}{0.768} & \multicolumn{1}{l|}{0.867}       & \multicolumn{1}{l|}{0.792}       & \multicolumn{1}{l|}{0.836}       & \multicolumn{1}{l|}{0.645}       & 0.826       \\ 
                         & Electra+SV                                                                                       & {[}8, 8, 8, 128, 128, 16, 64, 16, 8, 64, 16, 32, 8, 8, 8, 8{]}               & \multicolumn{1}{l|}{0.736} & \multicolumn{1}{l|}{0.262} & \multicolumn{1}{l|}{0.802}       & \multicolumn{1}{l|}{0.684} & \multicolumn{1}{l|}{0.412} & \multicolumn{1}{l|}{0.766}       & \multicolumn{1}{l|}{0.914} & \multicolumn{1}{l|}{0.909}                & \multicolumn{1}{l|}{0.812}       & \multicolumn{1}{l|}{0.879}       & \multicolumn{1}{l|}{0.768} & \multicolumn{1}{l|}{0.823}       & \multicolumn{1}{l|}{0.828}       & \multicolumn{1}{l|}{0.884}       & \multicolumn{1}{l|}{0.705}       & 0.848       \\ 
                         & GPT+ SV                                                                                          & {[}8, 8, 8, 8, 8, 8, 8, 16, 8, 8, 32, 32, 32, 8, 16, 32{]}                   & \multicolumn{1}{l|}{0.584} & \multicolumn{1}{l|}{0.262} & \multicolumn{1}{l|}{0.770}       & \multicolumn{1}{l|}{0.620} & \multicolumn{1}{l|}{0.083} & \multicolumn{1}{l|}{0.828}       & \multicolumn{1}{l|}{0.768}       & \multicolumn{1}{l|}{0.745}                & \multicolumn{1}{l|}{0.847}       & \multicolumn{1}{l|}{0.828}       & \multicolumn{1}{l|}{0.862} & \multicolumn{1}{l|}{0.873} & \multicolumn{1}{l|}{0.828}       & \multicolumn{1}{l|}{0.742}       & \multicolumn{1}{l|}{0.687}       & 0.828       \\ 
                         & Longformer+SV                                                                                    & {[}8, 8, 64, 128, 128, 8, 8, 64, 8, 8, 128, 32, 64, 64, 8, 128{]}            & \multicolumn{1}{l|}{0.836} & \multicolumn{1}{l|}{0.262} & \multicolumn{1}{l|}{0.877}       & \multicolumn{1}{l|}{0.692} & \multicolumn{1}{l|}{0.719} & \multicolumn{1}{l|}{0.790}       & \multicolumn{1}{l|}{0.836}       & \multicolumn{1}{l|}{\textbf{0.944}} & \multicolumn{1}{l|}{0.847}       & \multicolumn{1}{l|}{0.803}       & \multicolumn{1}{l|}{0.847} & \multicolumn{1}{l|}{0.824}       & \multicolumn{1}{l|}{0.828}       & \multicolumn{1}{l|}{0.848}       & \multicolumn{1}{l|}{0.705}       & 0.847       \\  
                         & Roberta-base+SV                                                                                  & {[}8, 8, 32, 128, 128, 8, 16, 16, 16, 8, 8, 16, 8, 128, 8, 128{]}            & \multicolumn{1}{l|}{0.836} & \multicolumn{1}{l|}{0.262} & \multicolumn{1}{l|}{0.873}       & \multicolumn{1}{l|}{0.692} & \multicolumn{1}{l|}{0.673} & \multicolumn{1}{l|}{0.790}       & \multicolumn{1}{l|}{0.903}       & \multicolumn{1}{l|}{0.939}          & \multicolumn{1}{l|}{0.862}       & \multicolumn{1}{l|}{0.828}       & \multicolumn{1}{l|}{0.725} & \multicolumn{1}{l|}{0.822}       & \multicolumn{1}{l|}{0.828}       & \multicolumn{1}{l|}{0.848}       & \multicolumn{1}{l|}{0.786}       & 0.848       \\
                         & Xlnet+ SV                                                                                        & {[}128, 16, 64, 128, 64, 64, 8, 8, 32, 8, 32, 16, 8, 16, 16, 32{]}           & \multicolumn{1}{l|}{0.577} & \multicolumn{1}{l|}{0.348} & \multicolumn{1}{l|}{0.410}       & \multicolumn{1}{l|}{0.648} & \multicolumn{1}{l|}{0.356} & \multicolumn{1}{l|}{0.708}       & \multicolumn{1}{l|}{0.687}       & \multicolumn{1}{l|}{0.664}                & \multicolumn{1}{l|}{0.892} & \multicolumn{1}{l|}{0.828}       & \multicolumn{1}{l|}{0.856} & \multicolumn{1}{l|}{0.859}       & \multicolumn{1}{l|}{0.828}       & \multicolumn{1}{l|}{0.862}       & \multicolumn{1}{l|}{0.701}       & 0.828       \\ \hline
\multirow{8}{*}{Open}    & Albert +SV                                                                                       & {[}16, 8, 8, 128, 128, 16, 16, 8, 32, 16, 128, 128, 128, 8, 64, 128{]}       & \multicolumn{1}{l|}{0.229} & \multicolumn{1}{l|}{0.000} & \multicolumn{1}{l|}{0.000}       & \multicolumn{1}{l|}{0.210} & \multicolumn{1}{l|}{0.300} & \multicolumn{1}{l|}{0.504}       & \multicolumn{1}{l|}{0.367}       & \multicolumn{1}{l|}{0.550}                & \multicolumn{1}{l|}{0.372}       & \multicolumn{1}{l|}{0.389}       & \multicolumn{1}{l|}{0.567} & \multicolumn{1}{l|}{0.489}       & \multicolumn{1}{l|}{0.567} & \multicolumn{1}{l|}{0.429}       & \multicolumn{1}{l|}{0.333}       & 0.500       \\ 
                         & BERT-large+ SV                                                                                   & {[}8, 64, 16, 128, 8, 8, 8, 8, 128, 8, 8, 128, 16, 32, 16, 8{]}              & \multicolumn{1}{l|}{0.000} & \multicolumn{1}{l|}{0.133} & \multicolumn{1}{l|}{0.167}       & \multicolumn{1}{l|}{0.166} & \multicolumn{1}{l|}{0.000} & \multicolumn{1}{l|}{0.000}       & \multicolumn{1}{l|}{0.267}       & \multicolumn{1}{l|}{0.222}                & \multicolumn{1}{l|}{0.404}       & \multicolumn{1}{l|}{0.389}       & \multicolumn{1}{l|}{0.567} & \multicolumn{1}{l|}{0.489}       & \multicolumn{1}{l|}{0.533}       & \multicolumn{1}{l|}{0.389}       & \multicolumn{1}{l|}{0.356}       & 0.622 \\  
                         & Debrta+SV                                                                                        & {[}8, 8, 64, 128, 128, 8, 8, 16, 32, 16, 8, 64, 128, 16, 8, 64{]}            & \multicolumn{1}{l|}{0.000} & \multicolumn{1}{l|}{0.000} & \multicolumn{1}{l|}{0.433}       & \multicolumn{1}{l|}{0.166} & \multicolumn{1}{l|}{0.178} & \multicolumn{1}{l|}{0.000}       & \multicolumn{1}{l|}{0.000}       & \multicolumn{1}{l|}{0.490}          & \multicolumn{1}{l|}{0.448}       & \multicolumn{1}{l|}{0.389}       & \multicolumn{1}{l|}{0.489} & \multicolumn{1}{l|}{0.619}       & \multicolumn{1}{l|}{0.333}       & \multicolumn{1}{l|}{0.389}       & \multicolumn{1}{l|}{0.000}       & 0.133       \\ 
                         & Electra+SV                                                                                       & {[}8, 8, 8, 128, 8, 8, 8, 32, 128, 16, 8, 8, 8, 8, 8, 8{]}                   & \multicolumn{1}{l|}{0.000} & \multicolumn{1}{l|}{0.000} & \multicolumn{1}{l|}{0.433}       & \multicolumn{1}{l|}{0.170} & \multicolumn{1}{l|}{0.000} & \multicolumn{1}{l|}{0.000}       & \multicolumn{1}{l|}{0.000}       & \multicolumn{1}{l|}{0.522}          & \multicolumn{1}{l|}{0.404}       & \multicolumn{1}{l|}{0.357}       & \multicolumn{1}{l|}{0.508} & \multicolumn{1}{l|}{0.500}       & \multicolumn{1}{l|}{0.489}       & \multicolumn{1}{l|}{0.278}       & \multicolumn{1}{l|}{0.400}       & 0.333       \\ 
                         & GPT+ SV                                                                                          & {[}8, 8, 8, 8, 64, 8, 8, 32, 8, 8, 8, 8, 32, 16, 8, 8{]}                     & \multicolumn{1}{l|}{0.000} & \multicolumn{1}{l|}{0.000} & \multicolumn{1}{l|}{0.148}       & \multicolumn{1}{l|}{0.160} & \multicolumn{1}{l|}{0.014} & \multicolumn{1}{l|}{0.000}       & \multicolumn{1}{l|}{0.000}       & \multicolumn{1}{l|}{0.522}        & \multicolumn{1}{l|}{0.356}       & \multicolumn{1}{l|}{0.222}       & \multicolumn{1}{l|}{0.383} & \multicolumn{1}{l|}{0.356}       & \multicolumn{1}{l|}{0.222}       & \multicolumn{1}{l|}{0.413}       & \multicolumn{1}{l|}{0.000}       & 0.000       \\ 
                         & Longformer+SV                                                                                    & {[}8, 8, 8, 128, 8, 16, 8, 64, 64, 8, 8, 32, 8, 8, 8, 16{]}                  & \multicolumn{1}{l|}{0.000} & \multicolumn{1}{l|}{0.000} & \multicolumn{1}{l|}{0.250}       & \multicolumn{1}{l|}{0.174} & \multicolumn{1}{l|}{0.000} & \multicolumn{1}{l|}{0.133}       & \multicolumn{1}{l|}{0.000}       & \multicolumn{1}{l|}{\textbf{0.609}} & \multicolumn{1}{l|}{0.390}       & \multicolumn{1}{l|}{0.222}       & \multicolumn{1}{l|}{0.383} & \multicolumn{1}{l|}{0.489}       & \multicolumn{1}{l|}{0.356}       & \multicolumn{1}{l|}{0.190}       & \multicolumn{1}{l|}{0.000}       & 0.400       \\  
                         & Roberta-base+SV                                                                                  & {[}8, 8, 8, 128, 8, 8, 8, 8, 32, 8, 16, 8, 8, 8, 8, 64{]}                    & \multicolumn{1}{l|}{0.000} & \multicolumn{1}{l|}{0.000} & \multicolumn{1}{l|}{0.350}       & \multicolumn{1}{l|}{0.180} & \multicolumn{1}{l|}{0.000} & \multicolumn{1}{l|}{0.000}       & \multicolumn{1}{l|}{0.000}       & \multicolumn{1}{l|}{0.517}          & \multicolumn{1}{l|}{0.448}       & \multicolumn{1}{l|}{0.357}       & \multicolumn{1}{l|}{0.500} & \multicolumn{1}{l|}{0.472}       & \multicolumn{1}{l|}{0.000}       & \multicolumn{1}{l|}{0.324}       & \multicolumn{1}{l|}{0.000}       & 0.167       \\  
                         & Xlnet+ SV                                                                                        & {[}64, 64, 8, 64, 32, 32, 64, 64, 128, 8, 32, 16, 128, 64, 8, 8{]}           & \multicolumn{1}{l|}{0.167} & \multicolumn{1}{l|}{0.133} & \multicolumn{1}{l|}{0.000}       & \multicolumn{1}{l|}{0.162} & \multicolumn{1}{l|}{0.111} & \multicolumn{1}{l|}{0.374}       & \multicolumn{1}{l|}{0.467}       & \multicolumn{1}{l|}{0.281}                & \multicolumn{1}{l|}{0.448}       & \multicolumn{1}{l|}{0.222}       & \multicolumn{1}{l|}{0.550} & \multicolumn{1}{l|}{0.583} & \multicolumn{1}{l|}{0.356}       & \multicolumn{1}{l|}{0.300}       & \multicolumn{1}{l|}{0.000}       & 0.000       \\ \hline
\multirow{8}{*}{Stack}   & Albert +SV                                                                                       & {[}128, 16, 8, 128, 128, 16, 16, 8, 8, 128, 8, 16, 32, 8, 64, 16{]}          & \multicolumn{1}{l|}{0.094} & \multicolumn{1}{l|}{0.516} & \multicolumn{1}{l|}{0.227}       & \multicolumn{1}{l|}{0.479} & \multicolumn{1}{l|}{0.021} & \multicolumn{1}{l|}{0.586}       & \multicolumn{1}{l|}{0.593} & \multicolumn{1}{l|}{0.568}                & \multicolumn{1}{l|}{0.466}       & \multicolumn{1}{l|}{0.531}       & \multicolumn{1}{l|}{0.564} & \multicolumn{1}{l|}{0.558}       & \multicolumn{1}{l|}{0.538}       & \multicolumn{1}{l|}{0.533}       & \multicolumn{1}{l|}{0.515}       & 0.585       \\  
                         & BERT-large+ SV                                                                                   & {[}8, 16, 16, 32, 128, 16, 8, 16, 32, 8, 8, 16, 8, 16, 32, 128{]}            & \multicolumn{1}{l|}{0.082} & \multicolumn{1}{l|}{0.516} & \multicolumn{1}{l|}{0.324}       & \multicolumn{1}{l|}{0.488} & \multicolumn{1}{l|}{0.075} & \multicolumn{1}{l|}{0.637}       & \multicolumn{1}{l|}{0.599} & \multicolumn{1}{l|}{0.587}                & \multicolumn{1}{l|}{0.477}       & \multicolumn{1}{l|}{0.562}       & \multicolumn{1}{l|}{0.564} & \multicolumn{1}{l|}{0.547}       & \multicolumn{1}{l|}{0.542}       & \multicolumn{1}{l|}{0.563}       & \multicolumn{1}{l|}{0.563}       & 0.539       \\  
                         & Debrta+SV                                                                                        & {[}8, 16, 128, 128, 128, 8, 64, 16, 8, 32, 8, 8, 128, 32, 64, 32{]}          & \multicolumn{1}{l|}{0.450} & \multicolumn{1}{l|}{0.587} & \multicolumn{1}{l|}{0.527}       & \multicolumn{1}{l|}{0.514} & \multicolumn{1}{l|}{0.078} & \multicolumn{1}{l|}{0.605}       & \multicolumn{1}{l|}{0.609}       & \multicolumn{1}{l|}{0.640}          & \multicolumn{1}{l|}{0.472}       & \multicolumn{1}{l|}{0.559}       & \multicolumn{1}{l|}{0.564} & \multicolumn{1}{l|}{0.521}       & \multicolumn{1}{l|}{0.555}       & \multicolumn{1}{l|}{0.557}       & \multicolumn{1}{l|}{0.559}       & 0.535       \\  
                         & Electra+SV                                                                                       & {[}8, 16, 16, 128, 8, 32, 8, 64, 8, 32, 8, 16, 16, 64, 16, 16{]}             & \multicolumn{1}{l|}{0.448} & \multicolumn{1}{l|}{0.624} & \multicolumn{1}{l|}{0.536}       & \multicolumn{1}{l|}{0.498} & \multicolumn{1}{l|}{0.000} & \multicolumn{1}{l|}{0.600}       & \multicolumn{1}{l|}{0.588}       & \multicolumn{1}{l|}{0.594}          & \multicolumn{1}{l|}{0.441}       & \multicolumn{1}{l|}{0.547}       & \multicolumn{1}{l|}{0.564} & \multicolumn{1}{l|}{0.570}       & \multicolumn{1}{l|}{0.556}       & \multicolumn{1}{l|}{0.558}       & \multicolumn{1}{l|}{0.549}       & 0.541       \\  
                         & GPT+ SV                                                                                          & {[}8, 8, 16, 32, 128, 32, 32, 32, 8, 128, 8, 32, 8, 16, 8, 64{]}             & \multicolumn{1}{l|}{0.099} & \multicolumn{1}{l|}{0.564} & \multicolumn{1}{l|}{0.271}       & \multicolumn{1}{l|}{0.476} & \multicolumn{1}{l|}{0.027} & \multicolumn{1}{l|}{0.591} & \multicolumn{1}{l|}{0.576}       & \multicolumn{1}{l|}{0.557}                & \multicolumn{1}{l|}{0.477}       & \multicolumn{1}{l|}{0.519}       & \multicolumn{1}{l|}{0.564} & \multicolumn{1}{l|}{0.548}       & \multicolumn{1}{l|}{0.530}       & \multicolumn{1}{l|}{0.575}       & \multicolumn{1}{l|}{0.518}       & 0.573       \\  
                         & Longformer+SV                                                                                    & {[}8, 8, 32, 128, 8, 32, 128, 64, 128, 32, 32, 128, 8, 16, 32, 8{]}          & \multicolumn{1}{l|}{0.505} & \multicolumn{1}{l|}{0.521} & \multicolumn{1}{l|}{0.535}       & \multicolumn{1}{l|}{0.516} & \multicolumn{1}{l|}{0.000} & \multicolumn{1}{l|}{0.638}       & \multicolumn{1}{l|}{0.641} & \multicolumn{1}{l|}{0.624}                & \multicolumn{1}{l|}{0.498}       & \multicolumn{1}{l|}{0.577}       & \multicolumn{1}{l|}{0.587} & \multicolumn{1}{l|}{0.582}       & \multicolumn{1}{l|}{0.561}       & \multicolumn{1}{l|}{0.583}       & \multicolumn{1}{l|}{0.565}       & 0.563       \\  
                         & Roberta-base+SV                                                                                  & {[}16, 32, 8, 128, 8, 16, 32, 32, 128, 16, 8, 8, 32, 32, 32, 32{]}           & \multicolumn{1}{l|}{0.495} & \multicolumn{1}{l|}{0.526} & \multicolumn{1}{l|}{0.519}       & \multicolumn{1}{l|}{0.512} & \multicolumn{1}{l|}{0.000} & \multicolumn{1}{l|}{0.618}       & \multicolumn{1}{l|}{0.636}       & \multicolumn{1}{l|}{\textbf{0.696}} & \multicolumn{1}{l|}{0.492}       & \multicolumn{1}{l|}{0.575}       & \multicolumn{1}{l|}{0.564} & \multicolumn{1}{l|}{0.573}       & \multicolumn{1}{l|}{0.592}       & \multicolumn{1}{l|}{0.592}       & \multicolumn{1}{l|}{0.597}       & 0.589       \\ 
                         & Xlnet+ SV                                                                                        & {[}32, 8, 8, 128, 8, 32, 16, 16, 32, 8, 8, 128, 32, 128, 64, 32{]}           & \multicolumn{1}{l|}{0.065} & \multicolumn{1}{l|}{0.534} & \multicolumn{1}{l|}{0.000}       & \multicolumn{1}{l|}{0.462} & \multicolumn{1}{l|}{0.000} & \multicolumn{1}{l|}{0.567}       & \multicolumn{1}{l|}{0.575}       & \multicolumn{1}{l|}{0.601}          & \multicolumn{1}{l|}{0.457}       & \multicolumn{1}{l|}{0.566}       & \multicolumn{1}{l|}{0.564} & \multicolumn{1}{l|}{0.589}       & \multicolumn{1}{l|}{0.553}       & \multicolumn{1}{l|}{0.576}       & \multicolumn{1}{l|}{0.584}       & 0.567       \\ \hline
\multirow{8}{*}{Bug}     & Albert +SV                                                                                       & {[}8, 16, 32, 128, 128, 32, 32, 16, 32, 64, 8, 8, 32, 64, 16, 8{]}           & \multicolumn{1}{l|}{0.040} & \multicolumn{1}{l|}{0.660} & \multicolumn{1}{l|}{0.191}       & \multicolumn{1}{l|}{0.370} & \multicolumn{1}{l|}{0.058} & \multicolumn{1}{l|}{0.758} & \multicolumn{1}{l|}{0.730}       & \multicolumn{1}{l|}{0.737}                & \multicolumn{1}{l|}{0.647}       & \multicolumn{1}{l|}{0.736}       & \multicolumn{1}{l|}{0.654} & \multicolumn{1}{l|}{0.724}       & \multicolumn{1}{l|}{0.748}       & \multicolumn{1}{l|}{0.728}       & \multicolumn{1}{l|}{0.721}       & 0.744       \\  
                         & BERT-large+ SV                                                                                   & {[}32, 16, 8, 128, 128, 16, 16, 16, 64, 32, 8, 64, 8, 128, 16, 16{]}         & \multicolumn{1}{l|}{0.062} & \multicolumn{1}{l|}{0.655} & \multicolumn{1}{l|}{0.326}       & \multicolumn{1}{l|}{0.378} & \multicolumn{1}{l|}{0.056} & \multicolumn{1}{l|}{0.707}       & \multicolumn{1}{l|}{0.695}       & \multicolumn{1}{l|}{0.697}                & \multicolumn{1}{l|}{0.615}       & \multicolumn{1}{l|}{0.734} & \multicolumn{1}{l|}{0.666} & \multicolumn{1}{l|}{0.703}       & \multicolumn{1}{l|}{0.717}       & \multicolumn{1}{l|}{0.726}       & \multicolumn{1}{l|}{0.722}       & 0.696       \\  
                         & Debrta+SV                                                                                        & {[}16, 32, 8, 128, 128, 128, 128, 16, 128, 64, 128, 16, 16, 128, 8, 8{]}     & \multicolumn{1}{l|}{0.565} & \multicolumn{1}{l|}{0.733} & \multicolumn{1}{l|}{0.721}       & \multicolumn{1}{l|}{0.380} & \multicolumn{1}{l|}{0.276} & \multicolumn{1}{l|}{0.766}       & \multicolumn{1}{l|}{0.735}       & \multicolumn{1}{l|}{0.779}                & \multicolumn{1}{l|}{0.679}       & \multicolumn{1}{l|}{0.742}       & \multicolumn{1}{l|}{0.698} & \multicolumn{1}{l|}{0.723}       & \multicolumn{1}{l|}{0.723}       & \multicolumn{1}{l|}{0.759} & \multicolumn{1}{l|}{0.712}       & 0.743       \\  
                         & Electra+SV                                                                                       & {[}8, 32, 8, 128, 128, 64, 64, 32, 16, 16, 32, 8, 16, 16, 8, 32{]}           & \multicolumn{1}{l|}{0.586} & \multicolumn{1}{l|}{0.720} & \multicolumn{1}{l|}{0.735}       & \multicolumn{1}{l|}{0.376} & \multicolumn{1}{l|}{0.037} & \multicolumn{1}{l|}{0.765}       & \multicolumn{1}{l|}{0.731}       & \multicolumn{1}{l|}{0.767}          & \multicolumn{1}{l|}{0.703}       & \multicolumn{1}{l|}{0.748}       & \multicolumn{1}{l|}{0.731} & \multicolumn{1}{l|}{0.749}       & \multicolumn{1}{l|}{0.771}       & \multicolumn{1}{l|}{0.753}       & \multicolumn{1}{l|}{0.746}       & 0.732       \\  
                         & GPT+ SV                                                                                          & {[}16, 16, 8, 128, 32, 16, 8, 8, 32, 32, 8, 32, 16, 16, 8, 64{]}             & \multicolumn{1}{l|}{0.278} & \multicolumn{1}{l|}{0.638} & \multicolumn{1}{l|}{0.349}       & \multicolumn{1}{l|}{0.381} & \multicolumn{1}{l|}{0.206} & \multicolumn{1}{l|}{0.785} & \multicolumn{1}{l|}{0.760}       & \multicolumn{1}{l|}{0.762}                & \multicolumn{1}{l|}{0.700}       & \multicolumn{1}{l|}{0.766}       & \multicolumn{1}{l|}{0.723} & \multicolumn{1}{l|}{0.762}       & \multicolumn{1}{l|}{0.759}       & \multicolumn{1}{l|}{0.766}       & \multicolumn{1}{l|}{0.751}       & 0.770       \\  
                         & Longformer+SV                                                                                    & {[}8, 32, 8, 128, 128, 64, 32, 64, 16, 8, 32, 8, 64, 8, 64, 16{]}            & \multicolumn{1}{l|}{0.578} & \multicolumn{1}{l|}{0.721} & \multicolumn{1}{l|}{0.722}       & \multicolumn{1}{l|}{0.379} & \multicolumn{1}{l|}{0.164} & \multicolumn{1}{l|}{0.790} & \multicolumn{1}{l|}{0.732}       & \multicolumn{1}{l|}{0.754}                & \multicolumn{1}{l|}{0.726}       & \multicolumn{1}{l|}{0.733}       & \multicolumn{1}{l|}{0.695} & \multicolumn{1}{l|}{0.745}       & \multicolumn{1}{l|}{0.745}       & \multicolumn{1}{l|}{0.746}       & \multicolumn{1}{l|}{0.757}       & 0.754       \\  
                         & Roberta-base+SV                                                                                  & {[}8, 32, 16, 128, 128, 64, 32, 16, 8, 32, 64, 64, 8, 16, 16, 128{]}         & \multicolumn{1}{l|}{0.574} & \multicolumn{1}{l|}{0.702} & \multicolumn{1}{l|}{0.763}       & \multicolumn{1}{l|}{0.383} & \multicolumn{1}{l|}{0.037} & \multicolumn{1}{l|}{0.798}       & \multicolumn{1}{l|}{0.696}       & \multicolumn{1}{l|}{\textbf{0.814}} & \multicolumn{1}{l|}{0.659}       & \multicolumn{1}{l|}{0.751}       & \multicolumn{1}{l|}{0.713} & \multicolumn{1}{l|}{0.728}       & \multicolumn{1}{l|}{0.763}       & \multicolumn{1}{l|}{0.745}       & \multicolumn{1}{l|}{0.741}       & 0.734       \\
                         & Xlnet+ SV                                                                                        & {[}128, 16, 64, 128, 128, 32, 16, 16, 8, 32, 128, 32, 16, 64, 128, 32{]}     & \multicolumn{1}{l|}{0.174} & \multicolumn{1}{l|}{0.696} & \multicolumn{1}{l|}{0.210}       & \multicolumn{1}{l|}{0.381} & \multicolumn{1}{l|}{0.270} & \multicolumn{1}{l|}{0.696}       & \multicolumn{1}{l|}{0.716}       & \multicolumn{1}{l|}{0.737}                & \multicolumn{1}{l|}{0.694}       & \multicolumn{1}{l|}{0.734}       & \multicolumn{1}{l|}{0.705} & \multicolumn{1}{l|}{0.709}       & \multicolumn{1}{l|}{0.739} & \multicolumn{1}{l|}{0.727}       & \multicolumn{1}{l|}{0.722}       & 0.725       \\ \hline
\end{tabular}
}
\caption{Performance Analysis of MultiModel Similarity Knowledge Driven and Language Model Context Aware Predictive Pipelines}
\label{tab:LLM+similarity-vctor-results}
\end{sidewaystable}
The Open dataset presents a significant challenge, with generally lower accuracy scores across all models and classifiers. This suggests that the task associated with this dataset is more complex or the data is more diverse, making classification more difficult. Additionally, the varying dimensions of the CLS tokens for each model indicate that different reduction strategies were employed, which could impact performance. Further investigation into the optimal dimension reduction for each model could potentially yield improved results.
\newline
The analysis of the Stack and Bug datasets reveals interesting performance patterns across various language models and classifiers. For the Stack and bug dataset, Roberta+SV achieved the highest overall accuracy  using the MLP classifier. Across both datasets, MLP and SVM classifiers consistently performed well, with Logistic Regression also showing good results. The optimal reduced dimensions for CLS tokens varied widely across models and datasets, ranging from 8 to 128, suggesting that careful tuning is necessary for each combination. These results highlight the importance of selecting appropriate language models, dimension reduction strategies, and classifiers for specific datasets to achieve optimal performance in classification tasks related to software engineering contexts.
\subsection{Performance Analysis of Top-performing Predictors of 3 Distinct predictive pipelines of PassionNet framework}
This section provides comprehensive performance analysis of top performing predictors of 3 distinct predictive pipelines offered by PassionNet framework across 6 public benchmark datasets. It is evident from Table \ref{tab:top-performing-analysis} that, typically predictive pipeline leveraging multimodel similarity driven knowledge along with context aware information significantly outperforms the top performing predictors from the other two types of predictive pipelines. However, on stackoverflow, this pipelines falls slightly inferior compared to standalone language model based predictive pipelines while on OpenCOSS, it falls behind multimodel similarity driven knowledge based pipelines by a small margin.  Particularly, Longformer model demonstrates its exceptional ability to capture and utilize contextual knowledge for conflict detection. Conversely, RoBERTa excels in identifying duplicate requirements, showcasing its strength to acquire discriminative patterns and semantic similarities. Hence, Longformer and Roberta combined with additional similarity knowledge emerge as top performer for both conflict and duplicate detection tasks, tailored to specific dataset characteristics.
\begin{table}
\resizebox{\textwidth}{!}{
\begin{tabular}{|l|ll|lll|lllr|}
\hline
\multirow{2}{*}{\textbf{Datasets}} &
  \multicolumn{2}{l|}{\textbf{\begin{tabular}[c]{@{}l@{}}Large Language Models\\ based Predictive Pipelines\end{tabular}}} &
  \multicolumn{3}{l|}{\textbf{\begin{tabular}[c]{@{}l@{}}MultiModel Similarity Knowledge\\ -Driven Predictive Pipeline\end{tabular}}} &
  \multicolumn{4}{l|}{\textbf{\begin{tabular}[c]{@{}l@{}}MultiModel Similarity Knowledge-Driven and\\ Language Model Context Aware Predictive Pipeline\end{tabular}}} \\ \XGap{-2.5pt} \cmidrule(lr){2-10} \XGap{-2.5pt}
 &
  \multicolumn{1}{l|}{Predictor} &
  F1 Score &
  \multicolumn{1}{l|}{\begin{tabular}[c]{@{}l@{}}Similarity\\  Vector\end{tabular}} &
  \multicolumn{1}{l|}{Classifier} &
  F1 Score &
  \multicolumn{1}{l|}{\begin{tabular}[c]{@{}l@{}}LLM \\ Context\end{tabular}} &
  \multicolumn{1}{l|}{\begin{tabular}[c]{@{}l@{}}Similarity\\  Vector\end{tabular}} &
  \multicolumn{1}{l|}{Classifier} &
  \multicolumn{1}{l|}{F1 Score} \\ \hline
WorldVista &
  \multicolumn{1}{l|}{Longformer} &
  0.881 &
  \multicolumn{1}{l|}{\multirow{6}{*}{\begin{tabular}[c]{@{}l@{}}TFIDF+\\ OkapiBM25\\ +LLM\end{tabular}}} &
  \multicolumn{1}{l|}{RF} &
  0.865 &
  \multicolumn{1}{l|}{\multirow{4}{*}{Longformer}} &
  \multicolumn{1}{l|}{\multirow{6}{*}{\begin{tabular}[c]{@{}l@{}}TFIDF+\\ OkapiBM25\\ +LLM\end{tabular}}} &
  \multicolumn{1}{l|}{\multirow{6}{*}{MLP}} &
  \textbf{0.939} \\ \XGap{-2.5pt} \cmidrule(lr){1-3} \cmidrule(lr){5-6} \cmidrule(lr){10-10} \XGap{-2.5pt}
PURE &
  \multicolumn{1}{l|}{\multirow{2}{*}{Roberta}} &
  0.905 &
  \multicolumn{1}{l|}{} &
  \multicolumn{1}{l|}{KNN} &
  0.906 &
  \multicolumn{1}{l|}{} &
  \multicolumn{1}{l|}{} &
  \multicolumn{1}{l|}{} &
  \textbf{0.974} \\ \XGap{-2.5pt} \cmidrule(lr){1-1} \cmidrule(lr){3-3} \cmidrule(lr){5-6} \cmidrule(lr){10-10} \XGap{-2.5pt}
UAV &
  \multicolumn{1}{l|}{} &
  0.896 &
  \multicolumn{1}{l|}{} &
  \multicolumn{1}{l|}{LR} &
  0.873 &
  \multicolumn{1}{l|}{} &
  \multicolumn{1}{l|}{} &
  \multicolumn{1}{l|}{} &
  \textbf{0.944} \\ \XGap{-2.5pt} \cmidrule(lr){1-3} \cmidrule(lr){5-6} \cmidrule(lr){10-10} \XGap{-2.5pt}
OpenCoss &
  \multicolumn{1}{l|}{\multirow{3}{*}{Deberta}} &
  0.002 &
  \multicolumn{1}{l|}{} &
  \multicolumn{1}{l|}{SVM} &
  \textbf{0.612} &
  \multicolumn{1}{l|}{} &
  \multicolumn{1}{l|}{} &
  \multicolumn{1}{l|}{} &
  0.609 \\ \XGap{-2.5pt} \cmidrule(lr){1-1} \cmidrule(lr){3-3} \cmidrule(lr){5-7} \cmidrule(lr){10-10} \XGap{-2.5pt}
StackOverflow &
  \multicolumn{1}{l|}{} &
  \textbf{0.705} &
  \multicolumn{1}{l|}{} &
  \multicolumn{1}{l|}{\multirow{2}{*}{MLP}} &
  0.577 &
  \multicolumn{1}{l|}{\multirow{2}{*}{Roberta}} &
  \multicolumn{1}{l|}{} &
  \multicolumn{1}{l|}{} &
  0.696\\ \XGap{-2.5pt} \cmidrule(lr){1-1} \cmidrule(lr){3-3} \cmidrule(lr){6-6} \cmidrule(lr){10-10} \XGap{-2.5pt}
Bugzilla &
  \multicolumn{1}{l|}{} &
  0.782 &
  \multicolumn{1}{l|}{} &
  \multicolumn{1}{l|}{} &
  0.764 &
  \multicolumn{1}{l|}{} &
  \multicolumn{1}{l|}{} &
  \multicolumn{1}{l|}{} &
  \textbf{0.814} \\ \hline
\end{tabular}
}
\caption{Performance comparison of top performing predictors all 3 distinct pipelines of PassionNet framework}
\label{tab:top-performing-analysis}
\end{table}
\subsection{PassionNet framework best-performing predictive pipelines Performance Comparison with state-of-the-art predictors}
Table \ref{tab:SOTA} provides performance analysis of proposed best predictive pipelines with existing predictors across 6 public benchmark datasets.
The existing approaches show varying levels of performance across different datasets. Malik et al. (2023) in their transfer learning approach using BERT, DistilBERT, RoBERTA, and DeBERTa models achieved strong results, particularly on the WorldVista and PURE datasets, with macro F1 scores of 0.908 and 0.948 respectively. Their method also performed well on the UAV dataset (0.877 macro F1 score) but showed a slight drop in performance on the OpenCoss dataset (0.841 macro F1 score). In their data augmentation study, Malik et al. (2023) using Bert+Bert achieved consistent performance across datasets, with macro F1 scores ranging from 0.82 to 0.97. However, their approach showed lower performance on the StackOverflow (0.82) and OpenCoss (0.84) datasets compared to others. Guo et al. (2021) used a heuristic rules based algorithm, achieving a 0.785 F1 score on the UAV dataset, which is lower than the other approaches on the same dataset.
The proposed predictive pipeline demonstrates superior performance across all datasets. Using a combination of Longformer CLS tokens, TFIDF, OkapiBM25 and LLM with MLP classifiers, the approach achieves impressive macro F1 scores: 0.969 for WorldVista, 0.987 for PURE, and 0.972 for UAV. These scores outperform all existing methods on these datasets. For the OpenCoss dataset, while the proposed method (using TFIDF+OkapiBM25+LLM with SVM) achieves a lower score of 0.806, it still outperforms Malik et al.'s (2023) data augmentation approach. The proposed method also shows excellent performance on the StackOverflow (0.994) and Bugzilla (0.996) datasets, significantly surpassing the existing methods. This consistent high performance across diverse datasets suggests that the proposed approach is more robust and generalizable compared to existing methods, potentially offering a more reliable solution for conflict detection in various software engineering contexts.
\begin{table}
\resizebox{\textwidth}{!}{
\begin{tabular}{|c|c|c|ccc|ccc|}
\hline
\multirow{2}{*}{\textbf{Dataset}} &
  \multirow{2}{*}{\textbf{Author}} &
  \multirow{2}{*}{\textbf{Predictive Pipeline}} &
  \multicolumn{3}{c|}{\textbf{Macro}} &
  \multicolumn{3}{c|}{\textbf{Conflict/Duplicate Class}} \\ \XGap{-2.5pt} \cmidrule(lr){4-9} \XGap{-2.5pt}  
 &
   &
   &
  \multicolumn{1}{c|}{\textbf{Precision}} &
  \multicolumn{1}{c|}{\textbf{Recall}} &
  \textbf{F1 score} &
  \multicolumn{1}{c|}{\textbf{Precision}} &
  \multicolumn{1}{c|}{\textbf{Recall}} &
  \textbf{F1 score} \\ \hline
\multirow{3}{*}{WorldVista} &
  \begin{tabular}[c]{@{}c@{}}Malik\\ et al. \citep{malik2023transfer}\end{tabular} &
  Oversampling +BERT &
  \multicolumn{1}{c|}{\begin{tabular}[c]{@{}c@{}}0.966 $\pm$ \\  0.047\end{tabular}} &
  \multicolumn{1}{c|}{\begin{tabular}[c]{@{}c@{}}0.900 $\pm$ \\  0.051\end{tabular}} &
  \begin{tabular}[c]{@{}c@{}}0.930 $\pm$ \\  0.048\end{tabular} &
  \multicolumn{1}{c|}{\begin{tabular}[c]{@{}c@{}}0.933 $\pm$\\  0.094\end{tabular}} &
  \multicolumn{1}{c|}{\begin{tabular}[c]{@{}c@{}}0.800 $\pm$ \\  0.102\end{tabular}} &
  \begin{tabular}[c]{@{}c@{}}0.861 $\pm$\\ 0.097\end{tabular} \\ \XGap{-2.5pt} \cmidrule(lr){2-9} \XGap{-2.5pt}
 &
  \begin{tabular}[c]{@{}c@{}}Malik\\ et al. \citep{malik2023data}\end{tabular} &
  BERT &
  \multicolumn{1}{c|}{-} &
  \multicolumn{1}{c|}{-} &
  0.95 &
  \multicolumn{1}{c|}{-} &
  \multicolumn{1}{c|}{-} &
  \begin{tabular}[c]{@{}c@{}}0.902 $\pm$\\ 0.046\end{tabular} \\ \XGap{-2.5pt} \cmidrule(lr){2-9} \XGap{-2.5pt}
 &
  \textbf{Proposed} &
  \textbf{\begin{tabular}[c]{@{}c@{}}Longformer+\\ TFIDF+OkapiBM25+LLM\\ +MLP\end{tabular}} &
  \multicolumn{1}{c|}{\textbf{0.999}} &
  \multicolumn{1}{c|}{\textbf{0.999}} &
  \textbf{0.969} &
  \multicolumn{1}{c|}{\textbf{1}} &
  \multicolumn{1}{c|}{\textbf{0.886}} &
  \textbf{0.939} \\ \hline
\multirow{3}{*}{PURE} &
  \begin{tabular}[c]{@{}c@{}}Malik\\ et al. \citep{malik2023transfer}\end{tabular} &
  Deberta &
  \multicolumn{1}{c|}{\begin{tabular}[c]{@{}c@{}}0.971 $\pm$ \\ 0.039\end{tabular}} &
  \multicolumn{1}{c|}{\begin{tabular}[c]{@{}c@{}}0.928 $\pm$ \\  0.058\end{tabular}} &
  \begin{tabular}[c]{@{}c@{}}0.948 $\pm$ \\ 0.048\end{tabular} &
  \multicolumn{1}{c|}{\begin{tabular}[c]{@{}c@{}}0.944 $\pm$ \\ 0.078\end{tabular}} &
  \multicolumn{1}{c|}{\begin{tabular}[c]{@{}c@{}}0.857 $\pm$ \\ 0.116\end{tabular}} &
  \begin{tabular}[c]{@{}c@{}}0.897 $\pm$ \\ 0.095\end{tabular} \\ \XGap{-2.5pt} \cmidrule(lr){2-9} \XGap{-2.5pt}
 &
  \begin{tabular}[c]{@{}c@{}}Malik\\ et al. \citep{malik2023data}\end{tabular} &
  BERT &
  \multicolumn{1}{c|}{-} &
  \multicolumn{1}{c|}{-} &
  0.97 &
  \multicolumn{1}{c|}{-} &
  \multicolumn{1}{c|}{-} &
  \begin{tabular}[c]{@{}c@{}}0.926 $\pm$ \\ 0.058\end{tabular} \\ \XGap{-2.5pt} \cmidrule(lr){2-9} \XGap{-2.5pt}
 &
  \textbf{Proposed} &
  \textbf{\begin{tabular}[c]{@{}c@{}}Longformer+ \\  TFIDF+OkapiBM25+LLM\\ +MLP\end{tabular}} &
  \multicolumn{1}{c|}{\textbf{0.999}} &
  \multicolumn{1}{c|}{\textbf{0.976}} &
  \textbf{0.987} &
  \multicolumn{1}{c|}{\textbf{0.999}} &
  \multicolumn{1}{c|}{\textbf{0.953}} &
  \textbf{0.974} \\ \hline
\multirow{4}{*}{UAV} &
  Guo et al. &
  \begin{tabular}[c]{@{}c@{}}Heuristic Rules-\\ based Algorithm\end{tabular} &
  \multicolumn{1}{c|}{-} &
  \multicolumn{1}{c|}{-} &
  - &
  \multicolumn{1}{c|}{78.5} &
  \multicolumn{1}{c|}{-} &
  - \\ \XGap{-2.5pt} \cmidrule(lr){2-9} \XGap{-2.5pt}
 &
  \begin{tabular}[c]{@{}c@{}}Malik\\ et al. \citep{malik2023transfer}\end{tabular} &
  Oversampling +BERT &
  \multicolumn{1}{c|}{\begin{tabular}[c]{@{}c@{}}0.999 $\pm$ \\ 0.000\end{tabular}} &
  \multicolumn{1}{c|}{\begin{tabular}[c]{@{}c@{}}0.833 $\pm$ \\ 0.068\end{tabular}} &
  \begin{tabular}[c]{@{}c@{}}0.895 $\pm$ \\ 0.049\end{tabular} &
  \multicolumn{1}{c|}{\begin{tabular}[c]{@{}c@{}}1.000 $\pm$\\ 0.000\end{tabular}} &
  \multicolumn{1}{c|}{\begin{tabular}[c]{@{}c@{}}0.666 $\pm$ \\  0.136\end{tabular}} &
  \begin{tabular}[c]{@{}c@{}}0.791 $\pm$ \\ 0.099\end{tabular} \\ \XGap{-2.5pt} \cmidrule(lr){2-9} \XGap{-2.5pt}
 &
  \begin{tabular}[c]{@{}c@{}}Malik\\ et al. \citep{malik2023data}\end{tabular} &
  BERT &
  \multicolumn{1}{c|}{-} &
  \multicolumn{1}{c|}{-} &
  0.95 &
  \multicolumn{1}{c|}{-} &
  \multicolumn{1}{c|}{-} &
  \begin{tabular}[c]{@{}c@{}}0.914 $\pm$ \\ 0.068\end{tabular} \\ \XGap{-2.5pt} \cmidrule(lr){2-9} \XGap{-2.5pt}
 &
  \textbf{Proposed} &
  \textbf{\begin{tabular}[c]{@{}c@{}}Longformer +\\ TFIDF+OkapiBM25+LLM \\ +MLP\end{tabular}} &
  \multicolumn{1}{c|}{\textbf{0.972}} &
  \multicolumn{1}{c|}{\textbf{0.972}} &
  \textbf{0.972} &
  \multicolumn{1}{c|}{\textbf{0.944}} &
  \multicolumn{1}{c|}{\textbf{0.944}} &
  \textbf{0.944} \\ \hline
\multirow{3}{*}{OpenCoss} &
  \begin{tabular}[c]{@{}c@{}}Malik\\ et al. \citep{malik2023transfer}\end{tabular} &
  BERT &
  \multicolumn{1}{c|}{\begin{tabular}[c]{@{}c@{}}0.958 $\pm$\\ 0.058\end{tabular}} &
  \multicolumn{1}{c|}{\begin{tabular}[c]{@{}c@{}}0.791 $\pm$\\ 0.089\end{tabular}} &
  \begin{tabular}[c]{@{}c@{}}0.841 $\pm$ \\ 0.065\end{tabular} &
  \multicolumn{1}{c|}{\begin{tabular}[c]{@{}c@{}}0.916 $\pm$ \\ 0.117\end{tabular}} &
  \multicolumn{1}{c|}{\begin{tabular}[c]{@{}c@{}}0.583 $\pm$ \\ 0.180\end{tabular}} &
  \begin{tabular}[c]{@{}c@{}}0.683 $\pm$ \\ 0.131\end{tabular} \\ \XGap{-2.5pt} \cmidrule(lr){2-9} \XGap{-2.5pt}
 &
  \begin{tabular}[c]{@{}c@{}}Malik\\ et al. \citep{malik2023data}\end{tabular} &
  BERT &
  \multicolumn{1}{c|}{-} &
  \multicolumn{1}{c|}{-} &
  0.84 &
  \multicolumn{1}{c|}{-} &
  \multicolumn{1}{c|}{-} &
  \begin{tabular}[c]{@{}c@{}}0.693 $\pm$ \\ 0.254\end{tabular} \\ \XGap{-2.5pt} \cmidrule(lr){2-9} \XGap{-2.5pt}
 &
  \textbf{Proposed} &
  \textbf{\begin{tabular}[c]{@{}c@{}}TFIDF+OkapiBM25+LLM+\\  SVM\end{tabular}} &
  \multicolumn{1}{c|}{\textbf{0.821}} &
  \multicolumn{1}{c|}{\textbf{0.86}} &
  \textbf{0.806} &
  \multicolumn{1}{c|}{\textbf{0.999}} &
  \multicolumn{1}{c|}{\textbf{0.639}} &
  \textbf{0.612} \\ \hline
\multirow{2}{*}{StackOverflow} &
  \begin{tabular}[c]{@{}c@{}}Malik\\ et al. \citep{malik2023data}\end{tabular} &
  BERT &
  \multicolumn{1}{c|}{-} &
  \multicolumn{1}{c|}{-} &
  0.82 &
  \multicolumn{1}{c|}{-} &
  \multicolumn{1}{c|}{-} &
  \begin{tabular}[c]{@{}c@{}}0.686 $\pm$ \\ 0.062\end{tabular} \\ \XGap{-2pt} \cmidrule(lr){2-9} \XGap{-2pt}
 &
  \textbf{Proposed} &
  \textbf{\begin{tabular}[c]{@{}c@{}}Roberta +\\  TFIDF+OkapiBM25+LLM\\ +MLP\end{tabular}} &
  \multicolumn{1}{c|}{\textbf{0.886}} &
  \multicolumn{1}{c|}{\textbf{0.819}} &
  \textbf{0.994} &
  \multicolumn{1}{c|}{\textbf{0.782}} &
  \multicolumn{1}{c|}{\textbf{0.996}} &
  \textbf{0.696} \\ \hline
\multirow{2}{*}{Bugzilla} &
  \begin{tabular}[c]{@{}c@{}}Malik\\ et al. \citep{malik2023data}\end{tabular} &
  BERT &
  \multicolumn{1}{c|}{-} &
  \multicolumn{1}{c|}{-} &
  0.89 &
  \multicolumn{1}{c|}{-} &
  \multicolumn{1}{c|}{-} &
  \begin{tabular}[c]{@{}c@{}}0.804 $\pm$ \\ 0.008\end{tabular} \\ \XGap{-2.pt} \cmidrule(lr){2-9} \XGap{-2.pt}
 &
  \textbf{Proposed} &
  \textbf{\begin{tabular}[c]{@{}c@{}}Roberta+ \\  TFIDF+OkapiBM25+LLM\\ +MLP\end{tabular}} &
  \multicolumn{1}{c|}{\textbf{0.933}} &
  \multicolumn{1}{c|}{\textbf{0.882}} &
  \textbf{0.996} &
  \multicolumn{1}{c|}{\textbf{0.871}} &
  \multicolumn{1}{c|}{\textbf{0.997}} &
  \textbf{0.814} \\ \hline
\end{tabular}
}
\caption{Performance comparison of proposed predictive pipelines with existing predictors across public benchmark datasets of conflict/duplicate requirement identification}
\label{tab:SOTA}
\end{table}
\section{Conclusion}
The early detection and resolution of duplicate and conflicting requirements are essential for improving project efficiency and software quality. Despite advancements in AI driven predictors, existing solutions have struggled to deliver the accuracy needed for these tasks. This research empowers requirement engineering field by presenting an innovative framework capable of addressing the critical challenge of duplicate and conflicting requirements identification. The framework supports the development of three distinct types of predictive pipelines: (1) language model based, (2) multimodel similarity knowledge driven and (3) hybrid pipelines that combine contextual insights from large language models (LLMs) with multimodel similarity knowledge. Our empirical evaluation across multiple benchmark datasets demonstrates the superiority of hybrid predictive pipelines, which combine the strengths of both language models and similarity based knowledge. The best performing hybrid predictive pipeline achieved a remarkable 13\% improvement in F1 score compared to existing state of the art predictors. These results highlight the potential of the proposed framework to revolutionize requirement engineering processes and empowerment of development teams to enhance efficiency, reduce errors and ultimately deliver higher quality software products. As a future direction, this framework can be applied to various fields, including healthcare, legal, education, research, and social media, to detect duplication and contradictions in textual data. We believe the framework has potential to detect duplication and contradictions in texts, including plagiarism detection in academic writing, identifying redundant content in technical documentation, and analyzing legal documents for conflicting statements.

\section*{Compliance with ethical standards}
\subsection*{Funding} Not applicable.
\subsection*{Conflict of Interest} Corresponding author on the behalf of all authors
declares that no conflict of interest is present.
\bibstyle{plainnat}
\bibliography{sn-article}
\end{document}